\documentclass[twocolumn,fleqn,natbib]{svjour2}
\usepackage{graphicx,url}
\usepackage{amsmath,amsfonts,amssymb,amsopn,amscd}
\usepackage{mathptmx}
\usepackage[centerlast]{subfigure}
\usepackage{color}
\usepackage{colortbl}
\usepackage{epsfig}

\newcommand{\be}{\begin{equation}}
\newcommand{\ee}{\end{equation}}
\newcommand{\bea}{\begin{eqnarray}}
\newcommand{\eea}{\end{eqnarray}}
\def\avg#1{\left\langle{#1}\right\rangle}	

\def\url#1{\textcolor{blue}{\underline{#1}}}	

\journalname{Journal of Computational Neuroscience}

\begin{document}



\title{Self-sustained asynchronous irregular states and Up--Down
states in thalamic, cortical and thalamocortical networks of
nonlinear integrate-and-fire neurons}


\author{Alain Destexhe}

\institute{Integrative and Computational Neuroscience Unit (UNIC), \\
UPR2191, CNRS, Gif-sur-Yvette, France \\
\email{Destexhe@unic.cnrs-gif.fr}}

\date{\today (in press)}

\maketitle



\begin{abstract}

Randomly-connected networks of integrate-and-fire (IF) neurons are
known to display asynchronous irregular (AI) activity states, which
resemble the discharge activity recorded in the cerebral cortex of
awake animals.  However, it is not clear whether such activity states
are specific to simple IF models, or if they also exist in networks
where neurons are endowed with complex intrinsic properties similar
to electrophysiological measurements.  Here, we investigate the
occurrence of AI states in networks of nonlinear IF neurons, such as
the adaptive exponential IF (Brette-Gerstner-Izhikevich) model.  This
model can display intrinsic properties such as low-threshold spike
(LTS), regular spiking (RS) or fast-spiking (FS).  We successively
investigate the oscillatory and AI dynamics of thalamic, cortical and
thalamocortical networks using such models.  AI states can be found
in each case, sometimes with surprisingly small network size of the
order of a few tens of neurons.  We show that the presence of LTS
neurons in cortex or in thalamus, explains the robust emergence of AI
states for relatively small network sizes.  Finally, we investigate
the role of spike-frequency adaptation (SFA).  In cortical networks
with strong SFA in RS cells, the AI state is transient, but when SFA
is reduced, AI states can be self-sustained for long times.  In
thalamocortical networks, AI states are found when the cortex is
itself in an AI state, but with strong SFA, the thalamocortical
network displays Up and Down state transitions, similar to
intracellular recordings during slow-wave sleep or anesthesia. 
Self-sustained Up and Down states could also be generated by
two-layer cortical networks with LTS cells.  These models suggest
that intrinsic properties such as adaptation and low-threshold
bursting activity are crucial for the genesis and control of AI
states in thalamocortical networks.

\end{abstract}




\section{Introduction}

In awake animals, the activity of single cortical neurons consist of
seemingly noisy activity, with very irregular discharges at
frequencies of 1-20~Hz and considerable fluctuations at the
level of the membrane potential (V$_m$) (Matsumara et al., 1988;
Steriade et al., 2001; Destexhe et al., 2003; Lee et al.,
2006).  Model networks of leaky integrate-and-fire (IF) neurons can
display activity states similar to the irregular spike discharge seen
in awake cortex.  These so-called ``asynchronous irregular'' (AI)
states contrast with the ``synchronous regular'' (SR) states, or with
oscillatory states (Brunel, 2000).  AI states have been observed more
recently as a self-sus\-tained activity in more realistic IF networks
with conduc\-tance-based synapses (Vogels and Abbott, 2005).  Such AI
states typically require large network sizes, of the order of a few
thousand neurons, to display characteristics consistent with
experimental data (El Boustani et al., 2007; Kumar et al., 2008).

In reality, neurons do not behave as leaky IF models, but rather
display complex intrinsic properties, such as adaptation or bursting,
and these intrinsic properties may be important for neuronal function
(Llinas, 1988).  However, it is not clear to what extent AI states
also appear in networks of more realistic neurons.  Similarly, the
genesis of AI states has never been investigated in the
thalamocortical system.  Recent efforts have been devoted to model
intrinsic neuronal properties using variants of the IF model (Smith
et al., 2000; Izhikevich, 2004; Brette and Gerstner, 2005).  In the
present paper, we use such models to analyze the genesis of AI states
in cortical, thalamic and thalamocortical networks of neurons
expressing complex intrinsic properties.


\section{Methods}

We successively describe the equations used for modeling neurons and
synapses, the connectivity of the different network models, as well
as the methods used to quantify network activity.

\subsection{Single-cell models}

To capture the intrinsic properties of central neurons, such as the
rebound bursting capabilities of thalamic cells and the
spike-frequency adaptation in cortex, we considered the
adaptive exponential IF (aeIF) model. This model consists of
the two-variable IF model proposed by Izhikevich (2004), which was
modified to include an exponential non-linearity around spike
threshold, based on the exponential IF model of Fourcaud-Trocme et
al. (2003).  These two models were combined by Brette and Gerstner
(2005), leading to the following set of equations:
{\small
\bea
  C_m {dV \over dt} & = & -g_L \ (V-E_L) \ + \ 
                     g_L \ \Delta \ \exp[(V-V_T)/\Delta] \ - \ w/S
                     \label{aeIF} \\ 
      {dw \over dt} & = & {1 \over \tau_w} \ [ a \ (V-E_L) - w ]
                     \nonumber ,
\eea
}
where $C_m$ = 1~$\mu$F/cm$^2$ is the specific membrane capacitance,
$g_L$ = 0.05~$mS/cm^2$ is the resting (leak) conductance, $E_L$ =
-60~mV is the resting potential (which is also equal to the reset
value after spike), $\Delta$ = 2.5~mV is the steepness of the
exponential approach to threshold, $V_T$ = -50~mV is the spike
threshold, and $S$ = 20,000~$\mu$m$^2$ is the membrane area.  When
$V$ reaches threshold, a spike is emitted and $V$ is instantaneously
reset and clamped to the reset value during a refractory period of
2.5~ms. $w$ is an adaptation variable, with time constant $\tau_w$ =
600~ms, and the dynamics of adaptation is given by parameter $a$ (in
$\mu$S).  At each spike, $w$ is incremented by a value $b$ (in nA),
which regulates the strength of adaptation, as analyzed in more
detail in the next section.

\subsection{Single-cell intrinsic properties}

\begin{figure}[h] 
\centerline{\psfig{figure=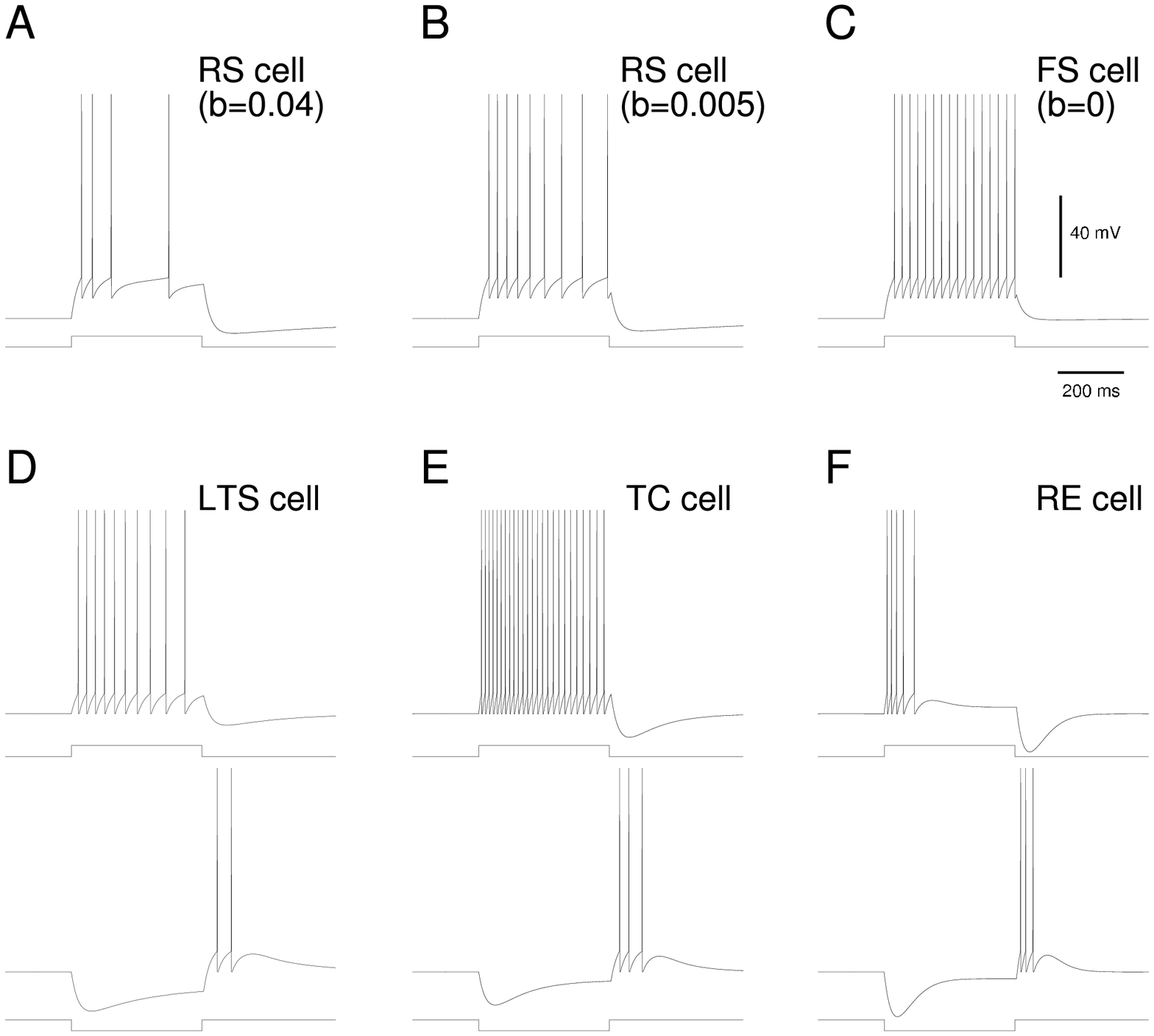,width=0.9\columnwidth}}

\caption{\label{intrinsic} Classes of neurons with different
intrinsic neuronal properties as modeled by the adaptive exponential
integrate-and-fire model.  A. Regular-spiking (RS) neuron with strong
adaptation.  B. RS neuron with weak adaptation.  C. Fast-spiking
(FS) cell with negligible adaptation.  D. Low-threshold spike (LTS)
cell.   E. Thalamocortical (TC) neuron.  F. Thalamic reticular (RE)
neuron.  In all cases, the response to a depolarizing current pulse
of 0.25~nA is shown on top.  For D-F, the bottom curves show the
response to a hyperpolarizing current pulse of -0.25~nA.
The units of parameter $b$ in A-C are nA.}

\end{figure}

The Brette-Gerstner-Izhikevich model was reported to display a wide
range of intrinsic neuronal properties (Izhikevich, 2004; Brette and
Gerstner, 2005).  We focus here only on a few cell types commonly
encountered in the thalamocortical system.  Cortical neurons were
modeled as ``regular spiking'' (RS) cells with spike-frequency
adaptation (Connors and Gutnick, 1990), which corresponds to the
parameters  $a$ = 0.001~$\mu$S and $b$ = 0.04~nA in the aeIF model
(Fig.~\ref{intrinsic}A).  The strength of adaptation can be modulated
by varying the parameter $b$, with $b$ = 0.005~nA for weakly adapting
cells (Fig.~\ref{intrinsic}B).  This parameter was estimated
heuristically based on Hodgkin-Huxley type models and RS cells found
in different preparations (Pospischil et al., 2008). For $b=0$, the
model generated responses with a negligible level of adaptation
(Fig.~\ref{intrinsic}C), similar to the ``fast-spiking'' (FS) cells
encountered in cortex, and which corresponds mostly to cortical
inhibitory neurons (Connors and Gutnick, 1990).  The latter will be
used to model cortical inhibitory interneurons in the present model.

Increasing the parameter $a$ leads to bursting activity (Izhikevich,
2004).  If we consider a moderate value of $a$ = 0.02~$\mu$S, the
model neuron also displays spike-frequency adaptation, even with
$b=0$ (Fig.~\ref{intrinsic}D, top).  However, this model also
generates a rebound burst in response to hyperpolarizing pulses,
while the conventional spike threshold is unchanged
(Fig.~\ref{intrinsic}D, bottom).  This behavior is similar to the
cortical low-threshold spike (LTS) cells (de la Pe\~na and
Geijo-Barrientos, 1996).

A further increase of parameter $a$ leads to more robust bursting
activity and weaker spike-frequency adaptation, producing patterns of
responses with moderate adaptation and strong rebound bursts
(Fig.~\ref{intrinsic}E).  This behavior was obtained for $a$ =
0.04~$\mu$S and $b=0$.  With larger values, $b$ = 0.08~$\mu$S and $a$
= 0.03~nA, the model generated bursting activity in response to both
depolarizing and hyperpolarizing stimuli (Fig.~\ref{intrinsic}F),
similar to thalamic reticular (RE) neurons (Destexhe and Sejnowski,
2003).  

Similar intrinsic properties as well as other types of behavior could
be generated by other combinations of parameters (see Izhikevich,
2004), but were not considered here.

\subsection{Network models}

Network models were constructed based on this aeIF model, according
to the following equations: 
{\small
\bea
  C_m {dV_i \over dt} & = & -g_L \ (V_i-E_L) \ + \ 
                     g_L \ \Delta_i \ \exp[(V-V_{Ti})/\Delta_i)
                     \ - \  w_i/S \nonumber \\
               & &   \ - \ \sum_j g_{ji} \ (V_i-E_j)
                     \label{aeIFnet} \\
      {dw_i \over dt} & = & {1 \over \tau_{w_i}} \
                     [ a_i \ (V_i-E_L) - w_i ]
                     \nonumber ,
\eea
}
where $V_i$ is the membrane potential of neuron $i$, and all
parameters are as in Eqs.~\ref{aeIF}, but were indexed to allow
variations according to cell type (see Results).  The term $\sum_j
g_{ji} \ (V_i-E_j)$ accounts for synaptic interactions, where
$g_{ji}$ is the conductance of the synapse from neuron $j$ to neuron
$i$ (and which can be zero), and $E_J$ is the reversal potential of
the synapse ($E_j$ = 0~mV for excitatory synapses and -80~mV for
inhibitory synapses).  Synaptic conductance time courses were
exponential; once the presynaptic cell fired, a fixed increment was
assigned to the corresponding $g_{ji}$ ($g_e$ and $g_i$ for
excitatory and inhibitory synapses, respectively), after which
$g_{ji}$ decays exponentially with a fixed time constant (5~ms for
excitation and 10~ms for inhibition).  Different synaptic
strengths $g_e$ and $g_i$ were considered depending on the network
type (see Results).  No synaptic delays were considered.

Note that because only small networks are considered here (of
the order of tens to a few thousand neurons), synaptic strengths need
necessarily to be large compared to physiological values.  Typical
values are $g_e$ = 6~nS and $g_i$ = 67~nS (Vogels and Abbott, 2005),
which correspond to postsynaptic potential sizes of 11~mV and 8.5~mV,
respectively, at -70~mV and at rest.  The amplitude of postsynaptic
potentials will be vastly different in an active network, with
typical values around 0.5~mV, respectively (El Boustani et al.,
2007).  Such a large difference is of course a property only seen in
conductance-based models.  It was shown that large networks (10,000
to over 100,000 neurons), with large numbers of synapses per neuron
($>500$), are necessary to achieve configurations with plausible
synaptic conductance values (El Boustani et al, 2007; Kumar et al.,
2008).

To initiate activity, a number of randomly-chosen neurons (from 2\%
to 10\% of the network) were stimulated by random excitatory inputs
during the first 50~ms of the simulation.  The mean frequency of this
random activity was high enough (200-400~Hz) to evoke random firing
in the recipient neurons.  In cases where self-sustained activity
appeared to be unstable, different parameters of this initial
stimulation were tested.  It is important to note that after
this initial period of 50~ms, no input was given to the network and
thus the activity states described here are self-sustained with no
external input or added noise.  The only source of noise was the
random connectivity (also termed ``quenched noise''.)

All equations were solved using the NEURON simulation environment
(Hines and Carnevale, 1997).

\subsection{Connectivity}

In all cases, the connectivity was random, but respected the
anatomical and morphological constraints about the connection
probability between the different cell types, as described below for
thalamus, cortex and thalamocortical relations.

\subsubsection{Thalamus}

The structure of the thalamus was approximated by a two layer network
of randomly connected cells, including one layer of thalamocortical
(TC) relay cells and one layer of thalamic reticular (RE) cells.  In
a number of species such as rodents, some thalamic nuclei are devoid
of interneurons (Jones, 1985).  There is also evidence that thalamic
interneurons do not play a major role in the genesis of internal
dynamics, for example in oscillations (Steriade et al., 1985; von
Krosigk et al., 1993).  Thus, thalamic interneurons were not
incorporated for simplicity.  The thalamic network had 10 times less
neurons compared to the cortical network, which corresponds to
morphological estimates (Sherman and Guillery, 2001).  Based on
anatomical data showing that axonal projections within the thalamic
circuitry are local but sparse (FitzGibbon et al., 1995; Jones, 1985;
Minderhoud, 1971), the excitatory projection from TC to RE cells had
a connection probability of 2\%, as in cortex, while the RE to TC
inhibitory projection was more dense (Kim et al., 1997), and was here
of a probability of about 8\%.  The same density was assumed from
inhibitory connections between RE cells.

This connectivity scheme corresponds to thalamic networks of size
$N$=100.  When comparing networks of different size, the connection
probability was rescaled inversely to network size, such that the
number of synapses received by the neurons was invariant to size.

\subsubsection{Cortex}

In area 5 of cat cerebral cortex, axon collaterals from pyramidal
cells are profuse and dense but remain localized within a few
hundreds of microns (Avenda\~no et al., 1988).  The connection
densities of cells in the cortical network were organized such that
each pyramidal cell (PY) or interneuron (IN) projected to a small
proportion of about 2\% of the size of the network.  The same
connection probability was also assumed for inhibitory connections. 
This connectivity was the same as that assumed in a previous model of
cortical AI states (Vogels and Abbott, 2005).  These connection
probabilities correspond to a cortical network of $N$=2000 neurons. 
As for thalamic networks, when different cortical network size were
compared, the connection probability was rescaled inversely to
network size to preserve the number of connections per neuron.

\subsubsection{Thalamocortical relations}

The thalamocortical and corticothalamic connections were also random,
and their densities were estimated from morphological studies as
follows.  Ascending thalamocortical fibers give most of their
synaptic contacts in layers I, IV and VI of cerebral cortex (White,
1986).  Given that layer VI pyramidal neurons constitute the major
source of corticothalamic fibers, these cells therefore mediate a
monosynaptic excitatory feedback loop (thalamus-cortex-thalamus;
White and Hersch, 1982), which was modeled here.  This monosynaptic
loop is also demonstrated by thalamically-evoked antidromic and
monosynaptic responses in the same, deeply lying cortical cell (see
Fig.~5 in Steriade et al., 1993b).  The model incorporated the fact
that all thalamic-projecting layer VI pyramidal cells connect TC
cells while leaving axon collaterals in the RE nucleus.  However,
lower layer V pyramids also project to thalamic nuclei, but they do
not leave collaterals in the RE nucleus (Bourassa and Desch\^enes,
1995); the latter were not modeled.  We did not include either the
influence of some thalamic intralaminar nuclei that project diffusely
to the cerebral cortex as well as receive projections from it (Jones,
1985).

Projections between thalamus and cortex are also local and sparse
(Avenda\~no et al., 1985; Jones, 1985; Robertson and Cunningham,
1981; Updyke, 1981) but have more divergence than intrathalamic or
intracortical connections (Bourassa and Desch\^enes, 1995; Freund et
al., 1989; Landry and Desch\^enes, 1981; Rausell and Jones, 1995). 
In the model, each PY cell had some probability of connecting to TC
and RE cells, but due to the large number of cortical neurons, this
corticothalamic connectivity was much more extended than local
thalamic connectivity, so that cortical synapses were majoritary in
thalamus (80\% of synapses in TC cells were from cortex), as
demonstrated experimentally (Sherman and Guillery, 2001).  Similarly,
each TC cell projected to PY and IN cells, using a connection
probability of 2\%.  In these conditions, cortical connectivity was
still dominated by intracortical synapses, as only about 6\% of
excitatory synapses were from thalamus, which corresponds well to
estimates from morphological estimates (Braitenberg and Shutz, 1998).

Thus, similar to a previous model (Destexhe et al., 1998), this
thalamocortical model can be thought of representing cerebral cortex
Layer~VI, connected reciprocally with its corresponding thalamic
area. All axonal projections of a given type were identical in extent
from cell to cell and all synaptic conductances were equal.  The
total synaptic conductance on each neuron was the same for cells of
the same type, independently of the size of the network.  The
connection probabilities are summarized in Table~1.

\newcommand{\ar}{\ $\rightarrow$\ }
\begin{table}[h]
 {\small
 \begin{tabular}{c|c|c}
  Connection type & Connection probability  & Nb.\ synapses / neuron \\
 \hline
  PY \ar PY  & 2 \%         & 32 \\
  PY \ar IN  & 2 \%         & 32 \\
  IN \ar PY  & 2 \%         & 8 \\
  IN \ar IN  & 2 \%         & 8 \\
  PY \ar TC  & 2 \%         & 32 \\
  PY \ar RE  & 2 \%         & 32 \\
  TC \ar RE  & 2 \%         & 2 \\
  RE \ar TC  & 8 \%         & 8 \\
  RE \ar RE  & 8 \%         & 8 
 \end{tabular}
 }

\caption{Connection probabilities between the different cell types. 
The probability is calculated for {\it outgoing} synapses, for
example for the TC \ar RE connection, the number indicated is the
probability that a given TC cell connects one RE cell.  In the last
column, an example of the average number of {\it incoming} synapses
(afferent synapses per neuron) is indicated for each type of
connection in a network with 1600 PY, 400 IN, 100 TC and 100 RE
neurons (TC = thalamocortical neurons; RE = thalamic reticular
neurons; PY = cortical excitatory neurons; IN = cortical inhibitory
interneurons).}

\end{table} 

\subsubsection{Two-layer cortical model}

Interlayer connectivity in cerebral cortex involves both
excitatory and inhibitory connections, but is predominantly
excitatory, with a connection density specific to the layers
considered (Thomson and Bannister, 2003; Binzegger et al., 2004). 
Interlayer (vertical) connectivity is also in general less dense than
intra-layer (horizontal) connections.  Two layers of cortical
networks were modeled as described above, with excitatory-only
interlayer connectivity with a probability of 1\%, which is twice
less dense than intra-layer connectivity (2\%; see Table~1).

\subsection{Quantification of network states}

Network states were quantified according to two aspects:
regularity and synchrony. To quantify the degree of temporal
regularity, we used the coefficient of variation (CV) of the
interspike intervals (ISI), averaged over all cells the network:
\be
  CV_{ISI} \ = \ \avg{ \frac{\sigma^{ISI}_i}{\overline{ISI}_i} } ~ ,
\ee
where the brackets $\avg{}$ indicate an average over all neurons,
while $\overline{ISI}_i$ and $\sigma^{ISI}_i$ are respectively the
mean and standard deviation of the ISIs of neuron $i$.  The
$CV_{ISI}$ is expected to take large values $\ge 1$ for temporally
irregular systems (the CV is 1 for a Poisson process).  In this
paper, we considered that a system is ``irregular'' if the $CV_{ISI}$
exceeds a value of 1.

The degree of synchrony was quantified using the averaged
pairwise cross-correlation between neurons in the network:
\be
  CC \ = \ \avg{ \frac{Cov(S_i,S_j)}{\sigma(S_i) \sigma(S_j)} } ~ ,
\ee
where the brackets $\avg{}$ indicate an average over a large number
of disjoint pairs of neurons (in general 500 pairs were used),
$Cov(S_i,S_j)$ is the covariance between two spike counts $S_i, S_j$,
and $\sigma(S_{i,j})$ is the standard deviation of each spike count. 
Spike counts were computed by counting spikes in successive time bins
of fixed duration (5~ms; all results were also checked using 2~ms). 
The $CC$ is comprised between -1 and 1, and takes high values only
for synchronous states.  A given network state can reasonably be
considered as ``asynchronous'' if $CC$ is low enough (typically $<
0.1$).  These criteria and methods were similar to that used
previously to characterize different states in network models
(Brunel, 2000; Kumar et al., 2008; El Boustani and Destexhe, 2009).
\\


\section{Results}

We successively consider network models for thalamus, cortex and the
thalamocortical system, and analyze their dynamics as a function of
their size and intrinsic properties.

\subsection{Networks of thalamic neurons}

\begin{figure}[h] 
\centerline{\psfig{figure=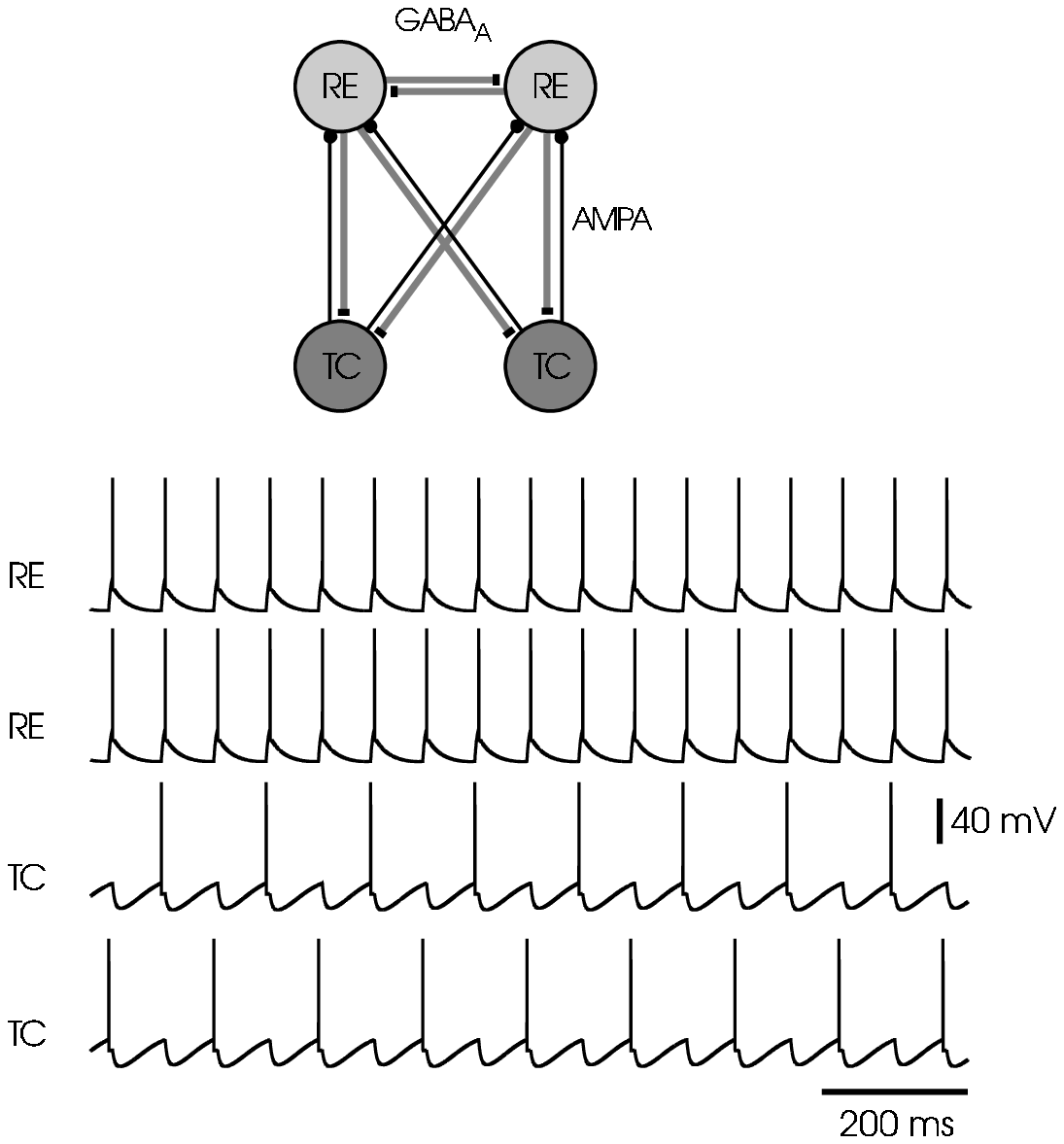,width=0.7\columnwidth}}

\caption{\label{spindle} Oscillatory behavior in simple circuits of
thalamic neurons.  Top scheme: minimal circuit for spindle
oscillations, consisting of two TC and two RE cells interconnected
with excitatory (AMPA) and inhibitory (GABA$_A$) synapses as
indicated.  This circuit generated oscillations at a frequency around
12~Hz, and in which RE cells fired at every cycle while TC cell fired
sub-harmonically, once every two cycles.  Synaptic conductance
values: $g_e$ = 30~nS, $g_i$ = 30~nS}

\end{figure}

Interconnected thalamic TC and RE cells can generate oscillations in
the spindle (7-14~Hz) frequency range (reviewed in Destexhe and
Sejnowski, 2003).  We first verified that the aeIF models outlined
above were capable of replicating this oscillatory behavior.  A small
circuit of interconnected TC and RE cells was built, where all cells
were interconnected, except for TC cells which do not have
interconnections (Fig.~\ref{spindle}, scheme).  As shown in
Fig.~\ref{spindle}, this circuit generated self-sustained
oscillations at a frequency around 10~Hz, with RE cells firing in
response to EPSPs from TC cells, and TC cells firing in rebound to
IPSPs from RE cells; the TC cells also generated subharmonic firing. 
These features are typical of spindle oscillations (Steriade, 2003). 
This behavior is only possible by taking into account the intrinsic
properties of thalamic cells, and in particular the rebound bursting
properties of TC cells (Destexhe and Sejnowski, 2003).  As in more
complex models, the oscillation frequency was dependent on the decay
kinetics of synaptic currents, and the oscillation was observed for a
large range of synaptic weights provided they were strong enough (not
shown).

\begin{figure}[h] 
\centerline{\psfig{figure=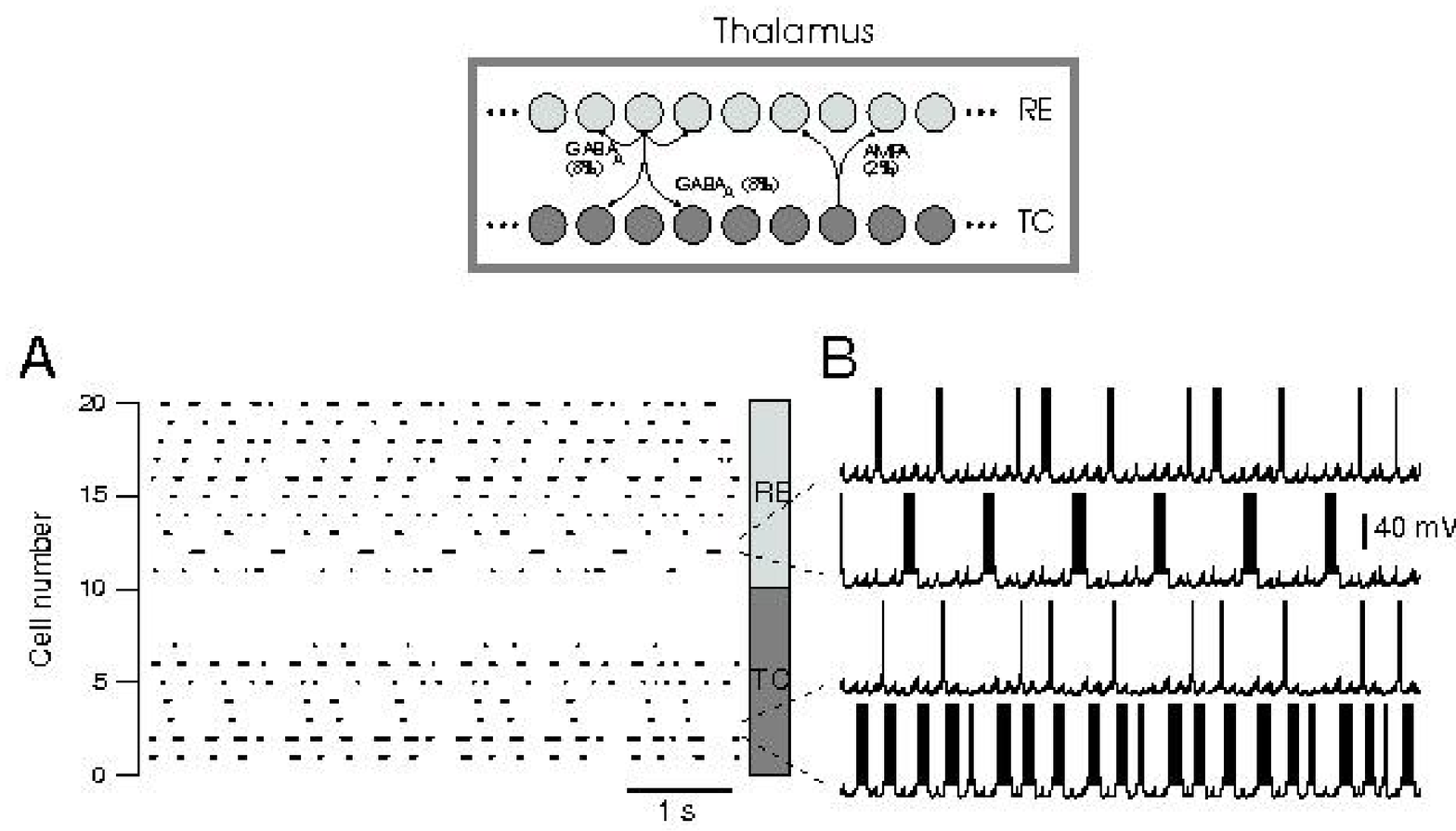,width=1.0\columnwidth}}

\caption{\label{thal20} Irregular oscillatory activity states in
networks of 20 randomly-connected thalamic neurons.  Top: scheme of
connectivity between the two layers of TC and RE cells, which were
randomly connected. A. Raster plot of the spiking activity of the 20
neurons.  B. V$_m$ activity of two cells of each type.  Synaptic
conductance values: $g_e$ = 6~nS, $g_i$ = 67~nS.  The connection
probability was of 8~\% (from RE to TC, and RE to RE) and 2\% (from
TC to RE), as in Table~1.  The activity is shown after a transient
time of 50~sec.  Inset in B: phase plot of the V$_m$ activity
of cell 1 against that of cell 2, showing the non-periodic character
of this activity (the $CV_{ISI}$ was of 1.36 and the $CC$ was of
0.025).}

\end{figure}

Small-size networks of TC and RE cells were considered by considering
more diffuse and random connectivity between the two TC and RE
layers, and weaker synaptic weights (see scheme in
Fig.~\ref{thal20}).  In these conditions, a $N$=20 network (10 TC and
10 RE cells) generated oscillatory behavior, but in contrast to the
small circuit considered above, these oscillations appeared to be
aperiodic, as seen from both rasterplot and single cells in
Fig.~\ref{thal20}. A phase plot between two TC cells
(Fig.~\ref{thal20}B, inset), as well as the high value of the
coefficient of variation ($CV_{ISI}$ = 1.36 in this case), confirmed
the aperiodic character of the oscillation. The raster plot also
shows that there is little synchrony in the firing of TC or RE cells,
unlike the circuit of Fig.~\ref{spindle}.  This was confirmed by the
low value of the averaged cross-correlation ($CC$ = 0.025).  This
state resembles the ``asynchronous regular'' states described earlier
(Brunel, 2000).  This behavior was robust to long integration times
(up to 100~sec were tested) without evidence for periodic behavior,
suggesting that this type of dynamics is self-sustained and
aperiodic.  Note that there is a possibility that these regimes are
periodic, but with a very long period which grows exponentially with
network size (Cessac, 2008; Cessac and Vi\'eville, 2009; see also
Crutchfield and Kaneko, 1988; T\'el and Lai, 2008).

\begin{figure}[h] 
\centerline{\psfig{figure=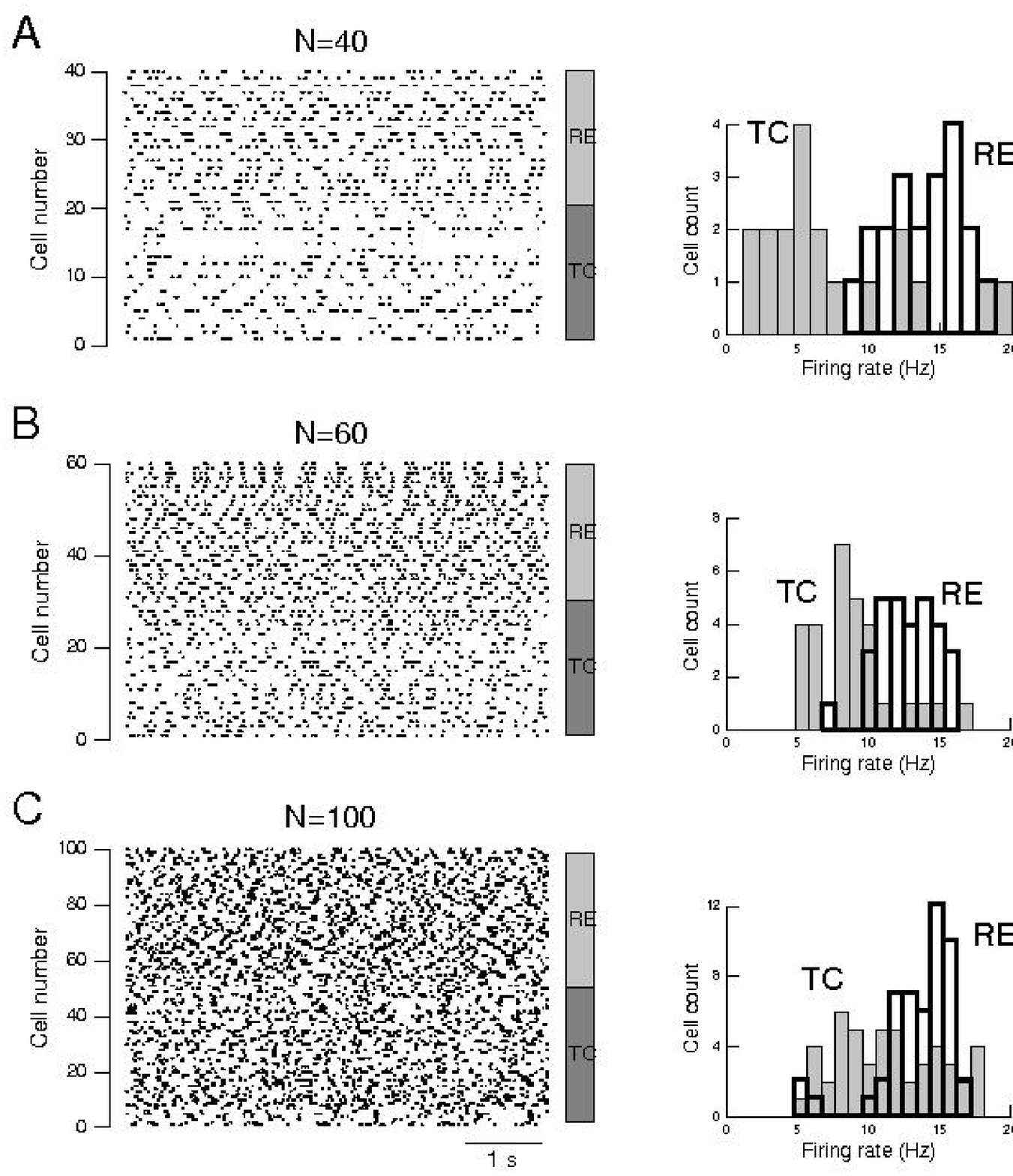,width=0.9\columnwidth}}

\caption{\label{thal40} Irregular activity states in thalamic
networks of various sizes.  The same connectivity scheme was used as
in Fig.~\ref{thal20}, but with larger sizes as indicated.  Same
synaptic conductance values and connection probabilities as in
Fig.~\ref{thal20}.  {\it Left panels:} activity after a
transient time of 50~sec. The $CV_{ISI}$ and $CC$ were respectively
of 1.45 and 0.027 for $N$=40, 1.35 and 0.02 for $N$=60, and 1.47 and
0.016 for $N$=100.  {\it Right panels:} distribution of firing
rates for the two cell types.}

\end{figure}

Networks comprising more thalamic neurons were considered based on a
similar connectivity scheme and the same synaptic weights as for
$N$=20.  By increasing the size from $N$=40 to $N$=100 generated
patterns of aperiodic oscillatory activity (Fig.~\ref{thal40}).  With
larger sizes, the activity became more and more similar to an
``asynchronous irregular'' state.  For example, for a $N$=100 network
(Fig.~\ref{thal40}C), the irregularity was high ($CV_{ISI}$ = 1.47)
and the synchrony was low ($CC = 0.016$).

\begin{figure}[h] 
\centerline{\psfig{figure=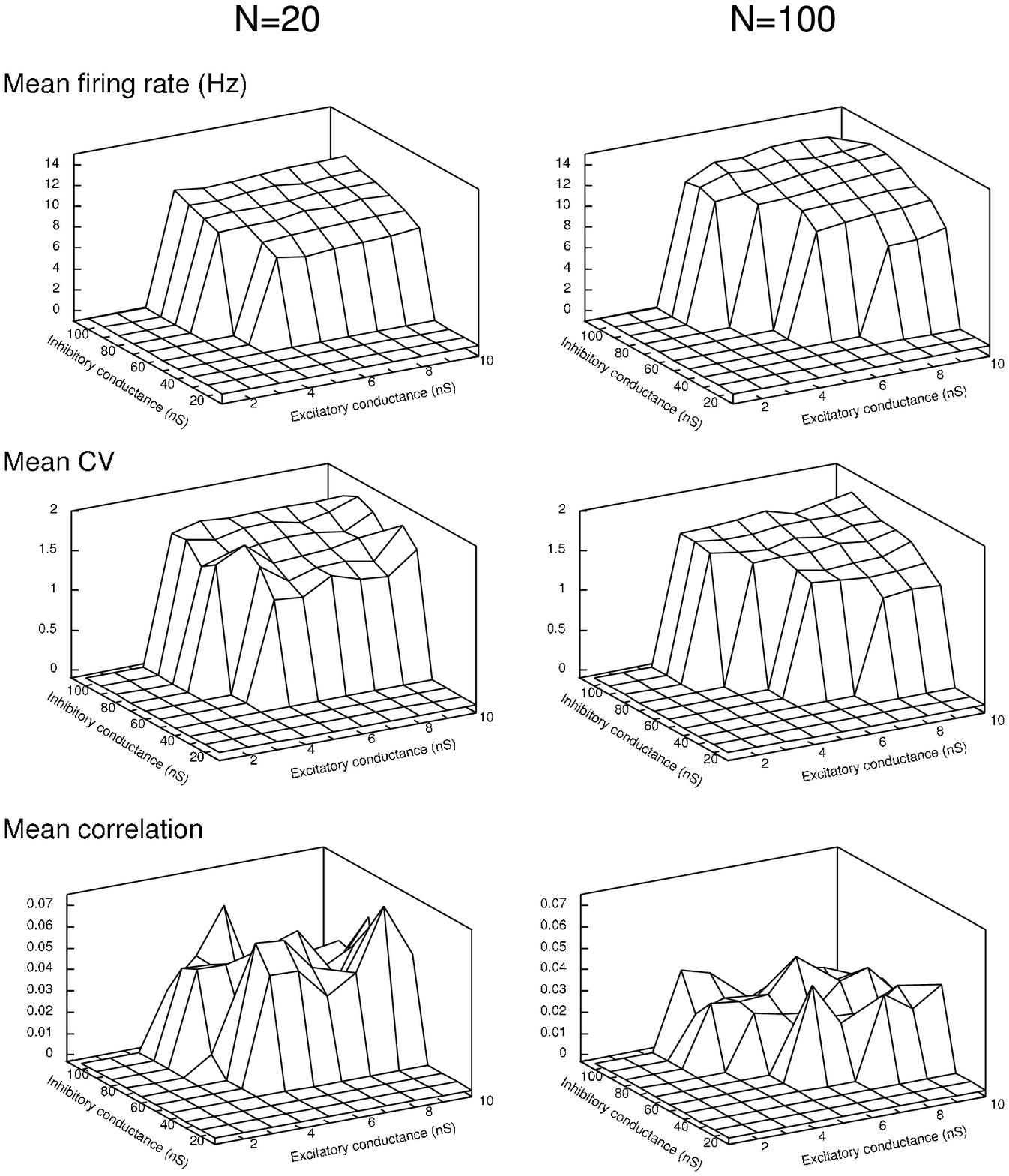,width=0.8\columnwidth}}

\caption{\label{thal3d} Domain of conductance parameters for
self-sustained irregular states in thalamic networks. {\it Top
Panels:} mean activity rate calculated over the entire network, shown
as a function of the synaptic conductance values ($g_e$ for
excitation and $g_i$ for inhibition) for $N$=20 (left) and $N$=100
(right).  {\it Middle panels: corresponding coefficients of variation
$CV_{ISI}$ calculated over all cells.  {\it Bottom panels:}
corresponding mean pairwise correlation $CC$ calculated over $N/2$
disjoint cell pairs in each network.} The irregular oscillations
appeared within a domain limited approximately by $g_e >$ 4~nS and
$g_i >$ 40~nS.  Transient oscillations that did not survive 10~sec
simulation time are not indicated.}

\end{figure}

This behavior was seen for a large domain of values for the synaptic
conductances.  This aspect, and more generally the robustness of
irregular oscillations to synaptic conductances, was quantified in
more detail in Fig.~\ref{thal3d}.  The domain of oscillatory behavior
was relatively large, provided both conductances were above some
threshold value (about 4~nS for excitatory conductance inputs and
40~nS for inhibitory conductances). The firing rate tended to
increase with the level of inhibition, which is presumably a
consequence of the rebound properties of thalamic neurons.  There
were also small differences between $N$=20 and $N$=100 networks: the
mean firing rate tended to be larger for $N=100$, while the averaged
pairwise correlation was lower (Fig.~\ref{thal3d}, bottom panels).

\subsection{Cortical networks}

\begin{figure}[h] 
\centerline{\psfig{figure=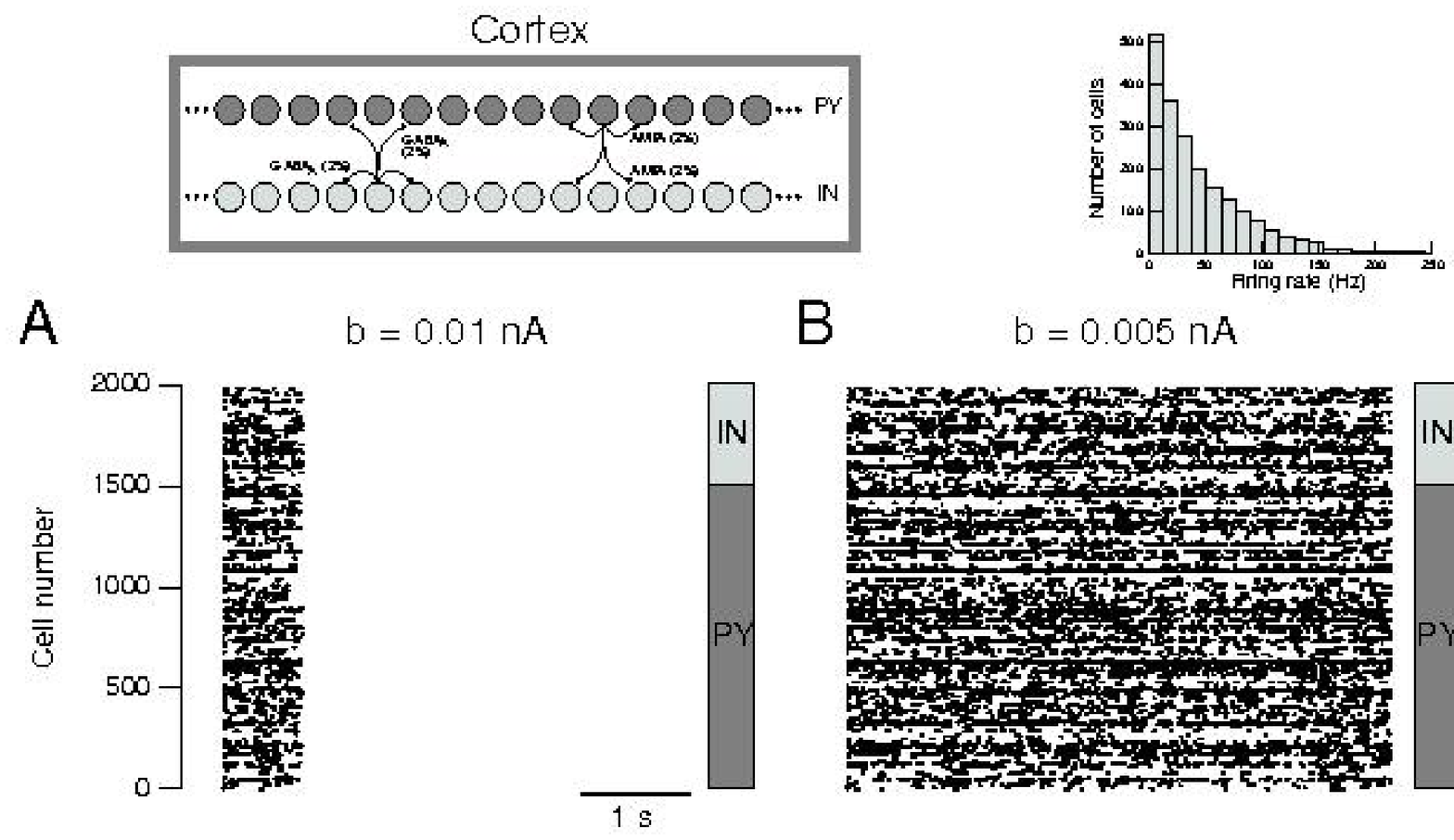,width=1.0\columnwidth}}

\caption{\label{cx2000} Transient and self-sustained irregular states
in randomly-connected networks of cortical neurons. Top: scheme of
connectivity.  A. Transient AI state in a $N$=2000 network, when the
RS neurons had a significant adaptation.  B.  Self-sustained AI state
in the same network after diminishing the strength of adaptation
($CV_{ISI}$ = 2.47, $CC$ = 0.005; Inset: distribution of firing
rates).   Synaptic conductances: $g_e$ = 6~nS, $g_i$ = 67~nS, and
connection probability as in Table~1.  Note that in this case,
PY cells are all of RS type, while all IN cells are of FS type.}

\end{figure}

We next considered the activity of cortical networks composed of
excitatory RS and inhibitory FS cells, randomly connected.  Unlike
thalamic networks, cortical circuits of this type do not generate
self-sustained AI states unless large networks are considered (Vogels
and Abbott, 2005; Kumar et al., 2008).  Compared to previous studies
using leaky IF models, the present model considers more complex IF
models, in particular for RS cells which display prominent
spike-frequency adaptation.  The cortical network had 80\% excitatory
and 20\% inhibitory cells, and was randomly connected (see scheme in
Fig.~\ref{cx2000}).  In these conditions, the genesis of
self-sustained AI states was possible, but was highly dependent on
the level of adaptation, as illustrated in Fig.~\ref{cx2000}.  For
strong to moderate adaptation, the network generated AI states but
they were transient and did not survive more than a few seconds
(Fig.~\ref{cx2000}A).  For weak adaptation, the AI states could be
sustained (Fig.~\ref{cx2000}B).

\begin{figure}[h] 
\centerline{\psfig{figure=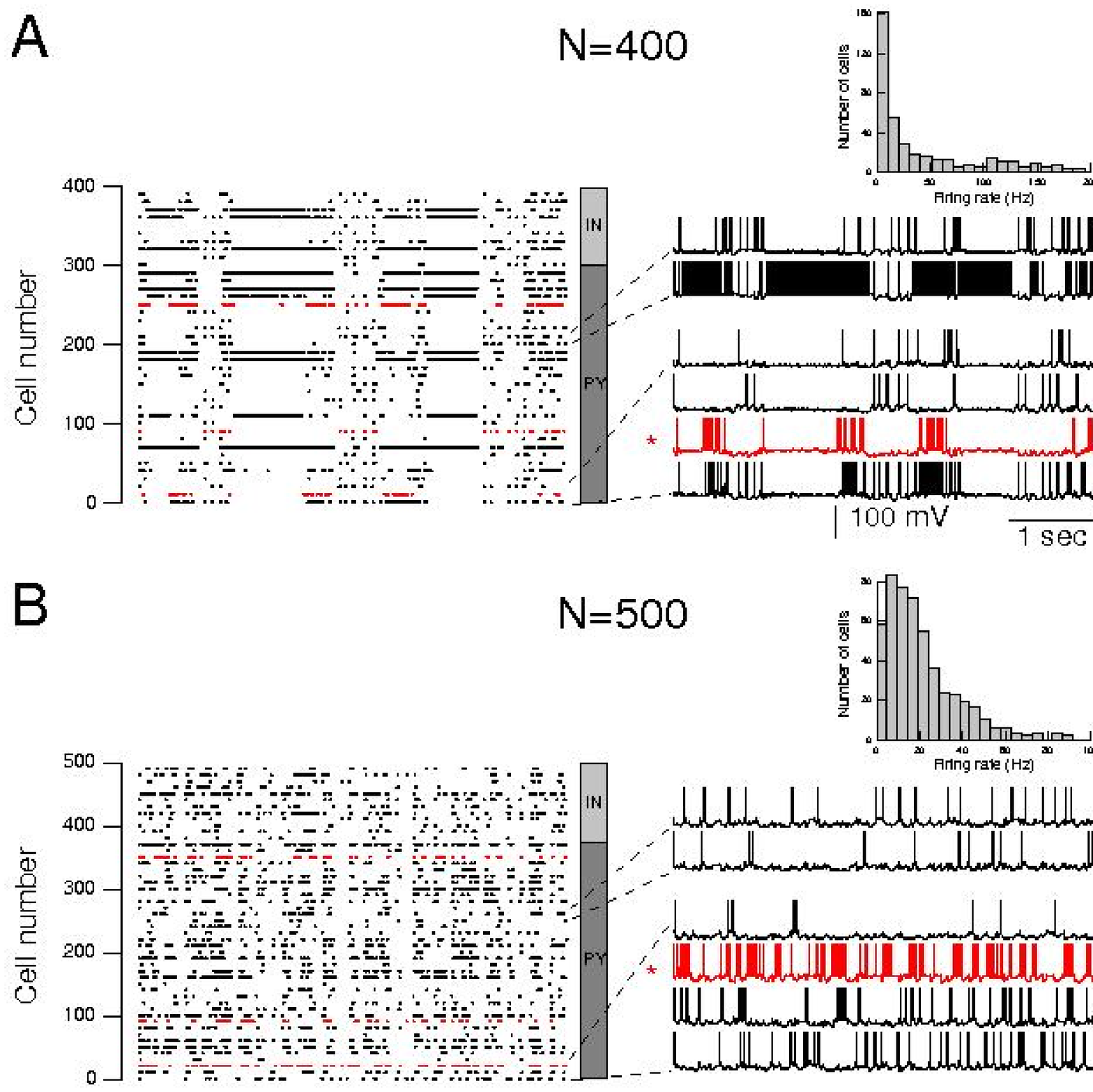,width=1.0\columnwidth}}

\caption{\label{cxLTS} Self-sustained irregular states in small
cortical networks with LTS cells.  The network was composed of
80\% excitatory (PY) cells and 20\% inhibitory (IN) cells.  95\% of
PY cells were RS type and 5\% were of LTS type, while all IN cells
were of FS type.  A. $N$=400 network displaying self-sustained
irregular activity ($CV_{ISI}$ = 2.84, $CC$ = 0.05).  B. AI state in
a $N$=500 network ($CV_{ISI}$ = 2.07, $CC$ = 0.05).  In each
case, the raster is shown on the left, with a few example cells on
the right, together with the firing rate distribution in insets.
Same synaptic conductances and connection probabilities as in
Fig.~\ref{cx2000}.  LTS cells are indicated in red (*).}

\end{figure}

The situation was radically different in the presence of LTS cells. 
Networks of relatively small size, $N$=400 (Fig.~\ref{cxLTS}A) or
$N$=500 (Fig.~\ref{cxLTS}B), could generate self-sustained AI states
if a proportion of LTS cells was present (LTS cells were connected
identically as for RS cells).  For $N$=400, the system
exhibited intermittent-like dynamics (Fig.~\ref{cxLTS}A), which was
typical of the transition between self-sustained oscillations, and AI
type behavior.  In this intermittency, the network switches between
SR and AI states, while for larger networks, the dynamics remain in
the AI state without intermittency (Fig.~\ref{cxLTS}B).

\begin{figure}[h] 
\centerline{\psfig{figure=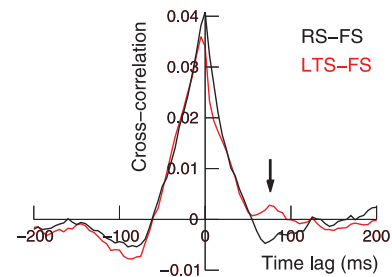,width=0.6\columnwidth}}

\caption{\label{cross} Sign of rebound activity in LTS cells
from cross-correlations.  Cross-correlations between RS and FS cells
(black), as well as between LTS and FS cells (red) were computed from
a $N=500$ network (same simulation as in Fig.~\ref{cxLTS}B, but with
a total time of 50~sec).  The cross-correlations were calculated from
instantaneous rates (spiking activity in successive 5~ms bins), and
were averaged between 400 non-overlapping pairs of cells for each
type.  The arrow shows that around 70~ms, a positive peak appears for
LTS-FS correlations, presumably due to the post-inhibitory rebound of
LTS cells.  As expected, all correlations peaked at negative delays,
betraying the excitatory action of RS and LTS cells onto FS cells.}

\end{figure}

To further characterize the role of LTS cells, the
cross-correl\-ation between LTS cells and inhibitory FS cells was
calculated and compared with the cross-correlation between RS and RS
cells (Fig.~\ref{cross}).  Because the firing FS cells leads only to
inhibition of RS cells, the cross-correlation is negative for
positive values of the time lag (Fig.~\ref{cross}, arrow in black
curve).  It is also negative for negative time lags, presumably due
to the inhibition between FS cells.  However, when correlating LTS
cells with FS cells, the pattern was different for positive time
lags, and positive correlations appeared (Fig.~\ref{cross}, arrow in
red curve).  This indicates that FS cells tend to excite LTS cells
after some delay of about 70~ms, which seems to correspond to the
post-inhibitory rebound because LTS and RS cells had the same
connectivity.  This positive value was also observed for other
network configurations with LTS cells (not shown).

\begin{figure}[h] 
\centerline{\psfig{figure=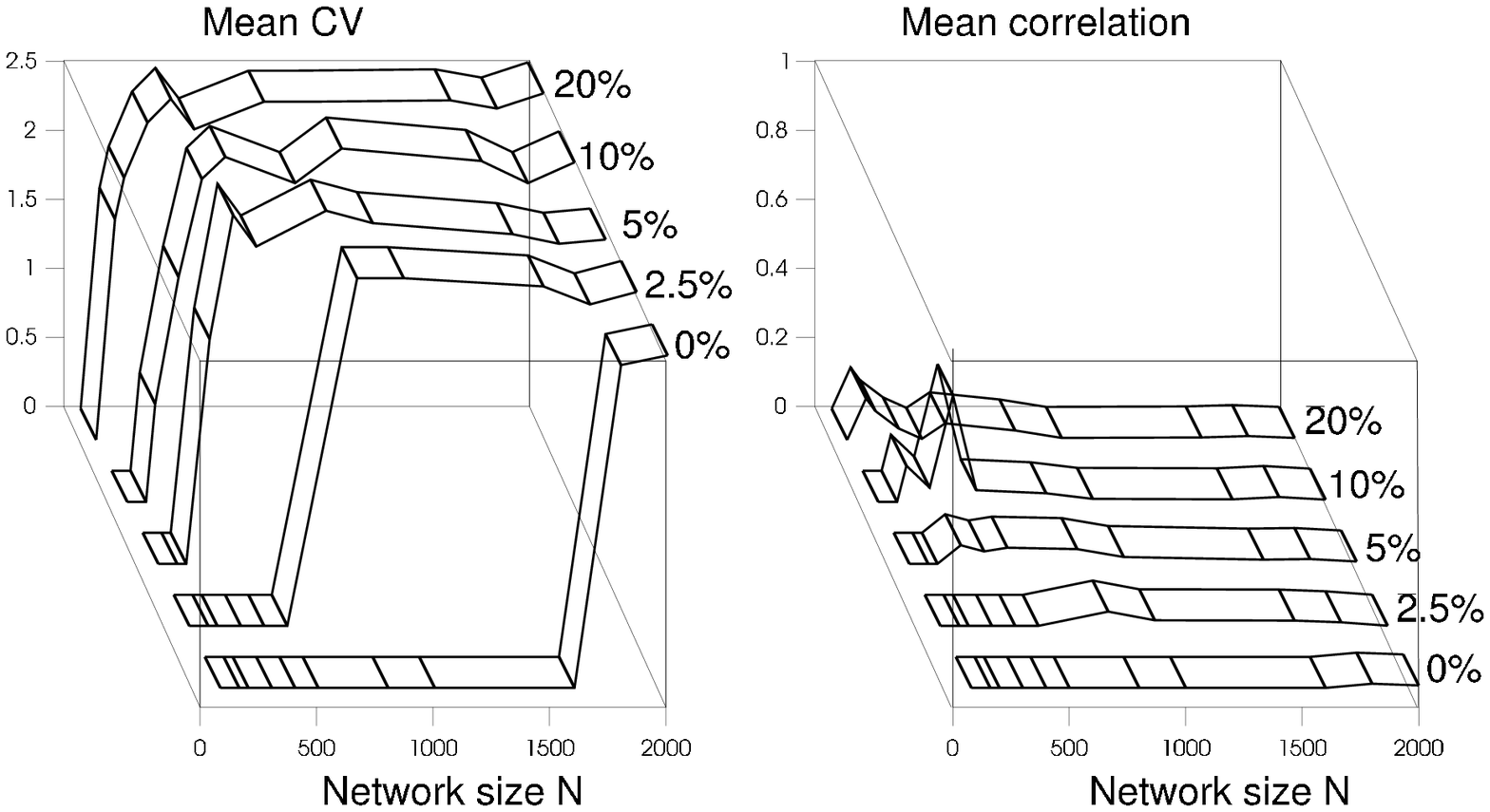,width=0.9\columnwidth}}

\caption{\label{cxsize} Minimal size for sustaining irregular states
in cortical networks.  Simulations similar to Fig.~\ref{cxLTS} were
performed for different network sizes (from 80 to 2000
neurons), and for different proportions of LTS cells (from 0
20
correlation coefficient ($CC$, right) are plotted as a function of
network size and proportion of LTS cells.  All self-sustained states
had high $CV$ and low $CC$, as indicated ($CV_{ISI}$ = 0 and $CC$ = 0
indicate states with no self-sustained activity.  The minimal size
for displaying AI states was much reduced if LTS cells were present. 
It was of 1800, 800, 300, 200 and 100 neurons, respectively for 0,
2.5, 5, 10 and 20\% of LTS cells.}

\end{figure}

These results suggest that, through their post-inhibitory rebound
property, the presence of LTS cells increases the excitability of the
network, enabling the genesis of AI states with relatively small
network sizes.  Indeed, by simulating different network size, with
and without LTS cells, revealed that the minimal size needed to
sustain AI states was highly reduced in the presence of LTS cells
(Fig.~\ref{cxsize}).  Moreover, the minimal size was inversely
related to the proportion of LTS cells (from 0\% to 20\%; see
Fig.~\ref{cxsize}).

\subsection{Thalamocortical networks}

\begin{figure}[h] 
\centerline{\psfig{figure=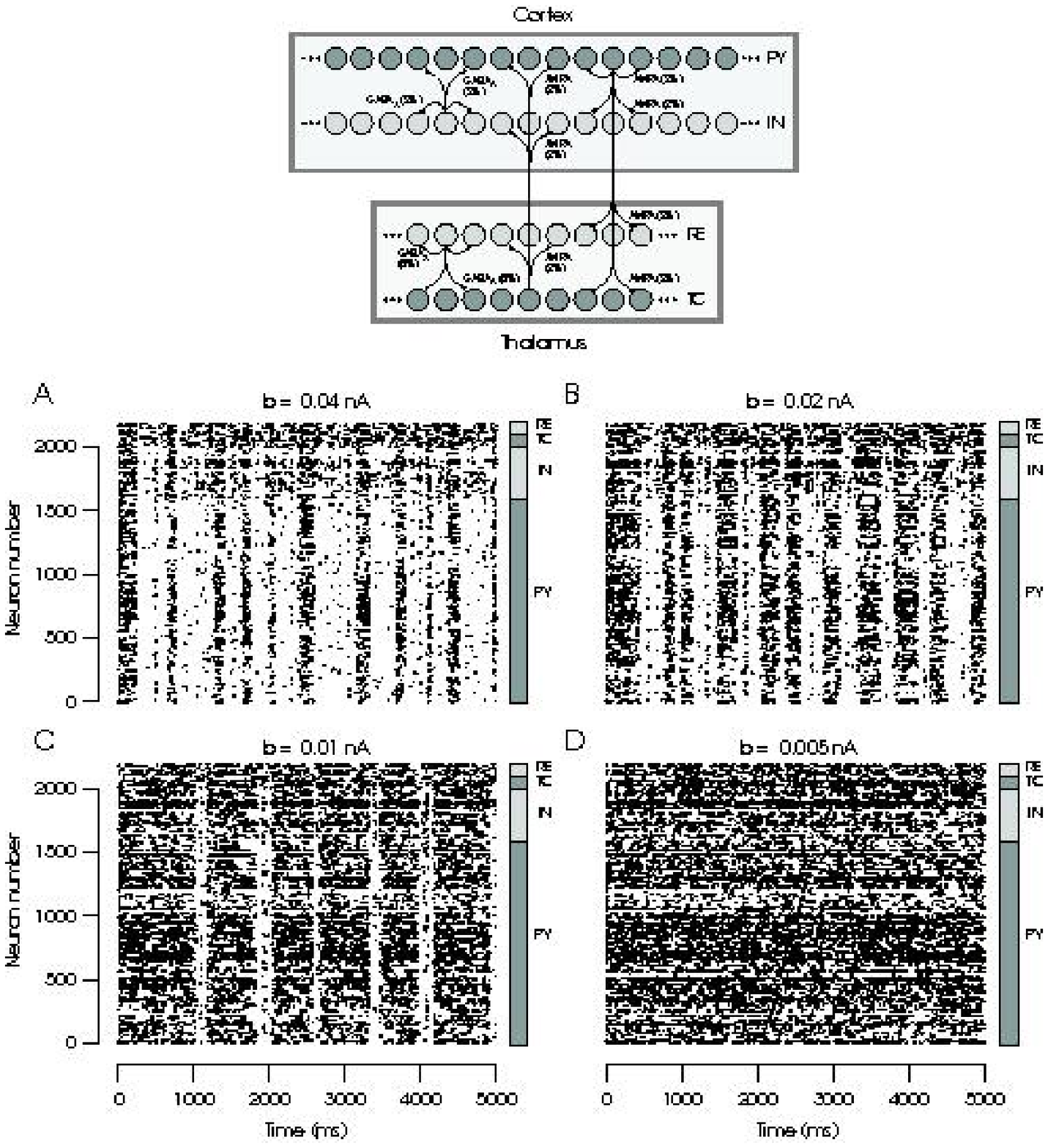,width=1.0\columnwidth}}

\caption{\label{thcsrasters} Self-sustained irregular and Up/Down
states in thalamocortical networks.  Top: Scheme of connectivity of
the thalamocortical network.  The network had 4 layers of cortical
pyramidal (PY), cortical interneurons (IN), thalamic reticular (RE)
and thalamocortical (TC) relay cells.  Each cell is represented by a
filled circle (dark gray = excitatory cells; light gray = inhibitory
cells), and synaptic connections are schematized by arrows.  Bottom
panels: From A to D, the same model was used (2200 cells total, 1600
PY, 400 IN, 100 TC and 100 RE cells), but with different strengths of
adaptation (from $b$=0.04~nA in A to $b$=0.005~nA in D). In all
rasters, only 10\% of cells are shown for each cell type, and the 4
layers of cells are indicated on the right. For the AI state in D,
cortical neurons were characterized by a mean firing rate of 44~Hz, a
coefficient of variation of $CV_{ISI}$ = 2.45 and a pairwise
correlation of $CC$ = 0.004.}

\end{figure}

\begin{figure}[h] 
\centerline{\psfig{figure=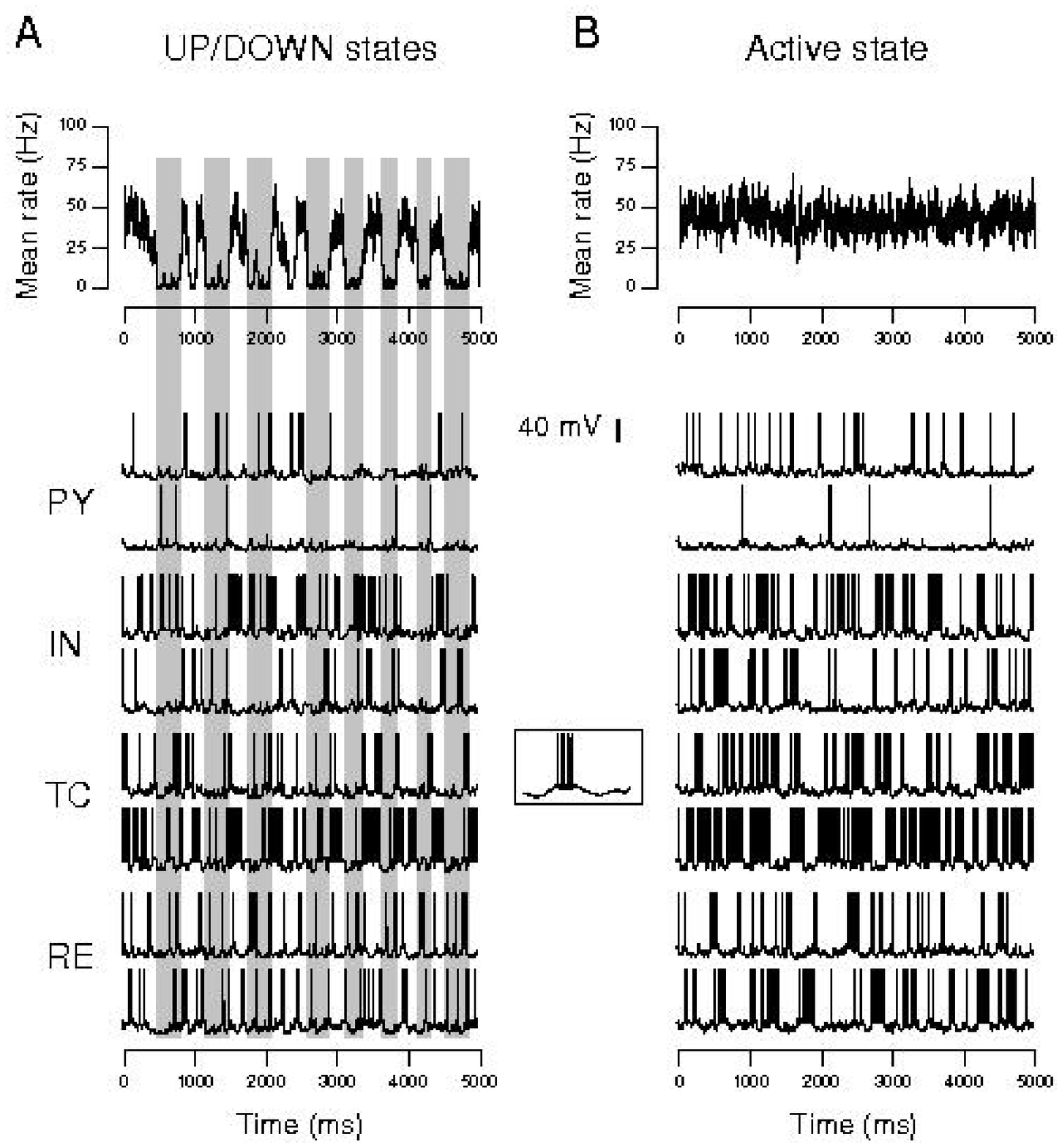,width=1.0\columnwidth}}

\caption{\label{thcx-updown} Cellular and averaged activity during
self-sustained states in thalamocortical networks.  The top graphs
indicate the instantaneous mean spike rate of cortical cells (in
successive 1~ms bins), while the bottom traces display the V$_m$
activity of two cells of each type as indicated.  A.  Up/Down state
transitions (same simulation as in Fig.~\ref{thcsrasters}B).  The
largest Down states are indicated in gray.  B.  Sustained active
state (same simulation as in Fig.~\ref{thcsrasters}D).  The same
cells are indicated for each simulation.  The inset shows a rebound
burst in a TC cell at 10 times higher temporal resolution.}

\end{figure}

The behavior of a thalamocortical network connected as schematized in
Fig.~\ref{thcsrasters} (top) also depends on the level of adaptation
in cortical cells.  With strong adaptation, the network displayed
alternating dynamics of active and silent periods (Up/Down states;
see Fig.~\ref{thcsrasters}A).  Progressively diminishing adaptation
(Fig.~\ref{thcsrasters}B-C) led to dynamics where the silent periods
(Down states) were diminished while the active (Up) states were
longer.  For weak adaptation, the network displayed self-sustained AI
states with no silent period (continuous Up state;
Fig.~\ref{thcsrasters}D). This latter state qualifies as an AI state,
with $CV_{ISI}$ = 2.45 and $CC$ = 0.004.  The corresponding cellular
activities are shown in Fig.~\ref{thcx-updown}.  Note that the
firing rate of the model (around 40~Hz on average) is larger than
experimental data, for both up states and AI states.  This is due to
the relatively small size of the network, as only networks of large
size can self-sustain AI states at low rates (El Boustani and
Destexhe, 2009).

\begin{figure}[h] 
\centerline{\psfig{figure=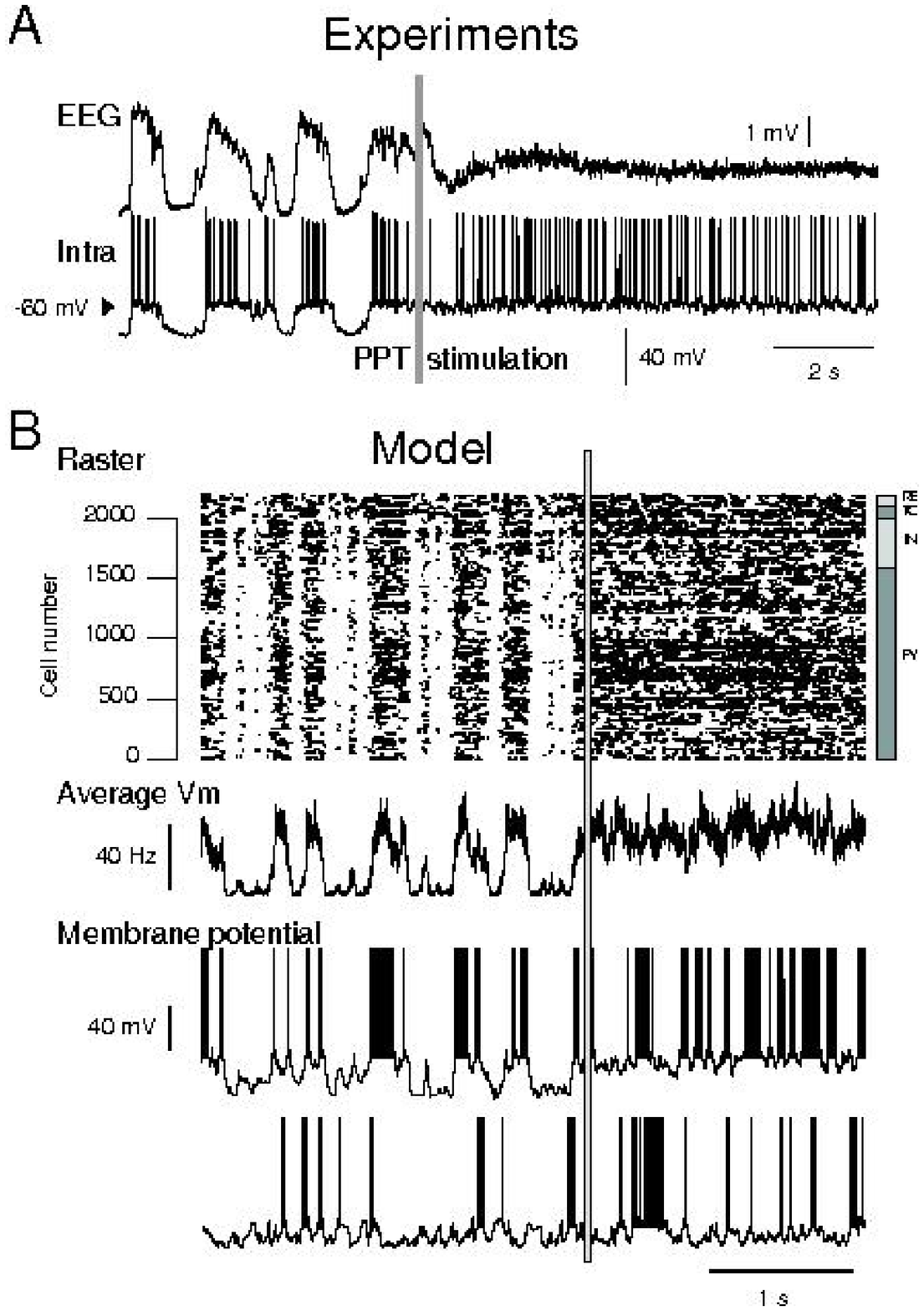,width=0.9\columnwidth}}

\caption{\label{ppt} Experiments and model of the transition from
Up/Down to activated states. A. Transition from Up/Down state
dynamics to an activated state, evoked by stimulation of the
pedonculo-pontine tegmentum (PPT) in an anesthetized cat.  The two
traces respectively show the EEG and intracellular activity recorded
in parietal cortex.  B. Similar transition obtained by changing the
value of $b$ from 0.02~nA to 0.005~nA (gray line).  All other
parameters were identical to Fig.~\ref{thcsrasters}.  Panel A
modified from Rudolph et al., 2005.}

\end{figure}

Because neuromodulatory substances, such as acetylcholine or
noradrenaline, block the K$^+$ conductances responsible for
adaptation (McCormick, 1992), the transition from Up \& Down states
to self-sustained active states in Fig.~\ref{thcsrasters} by reducing
spike-frequency adaptation reminds the brain activation process. 
Experimentally, a transition from Up/Down state dynamics to
self-sustained activated states can be obtained by electrical
stimulation of the brain stem, inducing a cascade of cholinergic
actions in thalamus and cortex (Fig.~\ref{ppt}A; Steriade et al.,
1993a; Rudolph et al., 2005). One of the most prominent of these
actions is a reduction in spike-frequency adaptation (McCormick,
1992).   A very similar transition can be obtained in the model when
the adaptation is reduced (from $b$=0.02~nA in A to $b$=0.005~nA; see
Fig.~\ref{ppt}B).

\subsection{Two-layer cortical networks}

\begin{figure}[h] 
\centerline{\psfig{figure=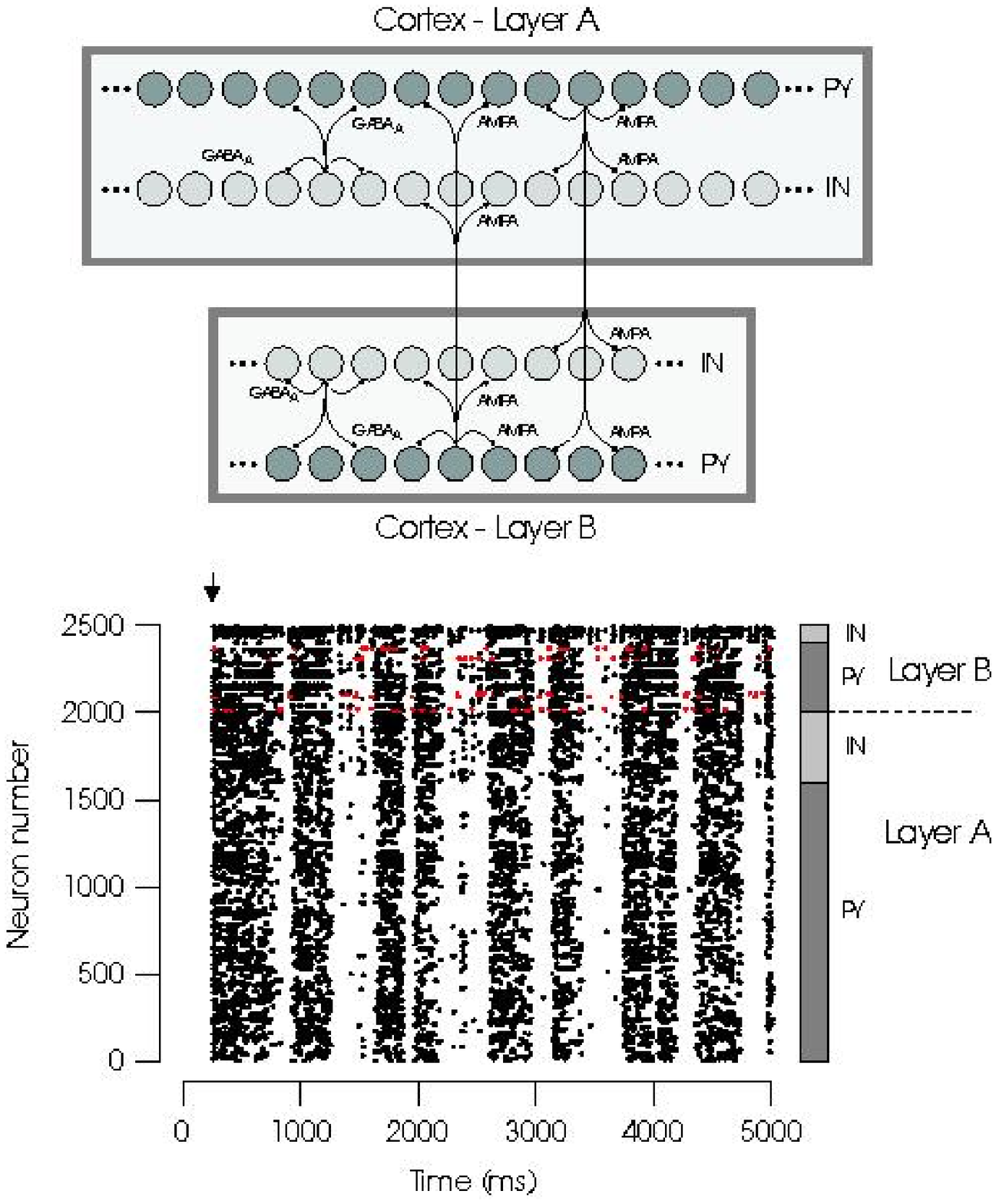,width=0.9\columnwidth}}

\caption{\label{2layers} Up/Down state dynamics in a 2-layer cortex
model with LTS cells.  {\it Top:} Scheme of connectivity between two
networks of $N$=2000 (Layer A) and $N$=500 (Layer B) neurons.  Layer
B had 10\% LTS cells and was capable of displaying self-sustained AI
states.  {\it Bottom:} Raster of the activity during 5 seconds (LTS
cells shown in red).  The stimulation of the network started at
$t$=250~ms (arrow), and switched the stable resting state to
self-sustained Up/Down state dynamics ($CV_{ISI}$ =  2.49, $CC$ =
0.069).  Parameters for the Layer~A network were identical to
Fig.~\ref{cx2000}A, and Layer~B was identical as Fig.~\ref{cxLTS}B,
with 10\% of LTS cells.  The interlayer connectivity was only
excitatory and had a connection probability of 1\%.}

\end{figure}

The above mechanism for Up/Down states relies on the fact that the
thalamus displays self-sustained AI states.  In the present section,
we present a similar mechanism but internal to cortex, in better
agreement with experiments which reported self-sustained Up/Down
state dynamics in cortical slices (Sanchez-Vives and McCormick, 2000;
Cossart et al., 2003).  To this end, the results from preceding
sections were combined into a ``two layer'' cortical network in which
one layer consisted of RS and FS cells, as described above
(Fig.~\ref{cx2000}A), while a second layer was a smaller network of
RS and FS cells, with some proportion of LTS cells, as in
Fig.~\ref{cxLTS}B.  These two networks were referred as ``Layer A''
and ``Layer B'', respectively (see scheme in Fig.~\ref{2layers}). 
When reciprocally connected with excitatory synapses, this 2-layer
system exhibited Up/Down state dynamics with a mechanism similar to
that described in the preceding section for thalamocortical networks.
In the present case, the Layer~B network generated self-sustained
activity, which served to ignite activity in the Layer~A (see raster
in Fig.~\ref{2layers}, bottom).  Interestingly, this simulation shows
that a network displaying transient dynamics (Fig.~\ref{cx2000}A)
connected with a network displaying AI states (Fig.~\ref{cxLTS}B)
yields a system displaying a different type of activity, Up and Down
states in this case.  This dynamics was entirely self-sustained and
coexisted with a stable resting state (Fig.~\ref{2layers}, arrow). 
All of these behaviors are indissociable from the particular
intrinsic properties of the neurons present in the system.  \\

\section{Discussion}

In this paper, we have shown that networks of neurons displaying
complex intrinsic properties can display various type of AI states. 
The main finding were that (a) thalamic networks, where neurons are
endowed with rebound bursting capabilities, can display AI states for
remarkably small size (N$\sim$100). (b) Cortical networks display AI
states, as reported previously (Brunel 2000, Vogels and Abbott,
2005), but when adaptation is included in excitatory neurons,
cortical networks generate AI states only for weak adaptation.  (c)
Including a small proportion of LTS cells in cortical networks
greatly reduces the minimal size needed to generate AI states.  (d)
thalamocortical networks can display AI states or Up/Down state
dynamics, depending on the level of adaptation in cortical cells. 
(e) Two-layer cortical networks can self-sustain Up/Down states
solely from intrinsic dynamics.

These numerical observations suggest that spike-frequency adaptation
acts against the genesis of AI states, and tends to silence the
network (Fig.~\ref{cx2000}).  Adaptation diminishes excitability, and
several neuromodulators increase cortical excitability by blocking
the slow K$^+$ conductances responsible for adaptation (McCormick,
1992).  Neuromodulation has also strong effects on leak conductances,
in particular in the thalamus (McCormick, 1992).  We did not attempt
to incorporate such effects in this model, but the precise modeling
of neuromodulation, and possible transitions from AI to various types
of oscillatory behavior, constitutes a possible extension of the
model.

The presence of LTS neurons tends to greatly favor AI states.  This
is consistent with the genesis of AI states in very small thalamic
networks (N$\sim$100; Fig.~\ref{thal40}), where all neurons display
LTS.  In cortex, inclusion of a small proportion of LTS cells (as
observed experimentally; see de la Pena and Geijo-Barrientos, 1996;
Destexhe et al., 2001), greatly reduces the minimal size to display
AI states.  The probable mechanisms is that LTS cells renders the
network insensitive to ``gaps'' of firing, caused by occasional
synchronized inhibition, and which usually stops the activity.  These
gaps are followed by rebound bursts in a minority of LTS cells if
they are present, and thus renders the network less vulnerable to
such gaps.  This also suggests that LTS cells could be indicative of
networks that generate AI type of activity.  It would be interesting
to investigate in more detail, and perhaps theoretically, the
contrasting effects of adaptation and LTS on the genesis of AI
states.  Inclusion of other types of bursting cells, such as
intrinsically bursting neurons (Connors and Gutnick, 1990) or
inhibitory LTS cells (Xiang et al., 1998), also constitutes a
possible extension of this study.

In the thalamocortical system, the association of thalamic networks
and cortical networks generates a variety of states, including AI
states and different forms of Up/Down state dynamics, for different
levels of adaptation in cortical excitatory cells.  Reducing
adaptation, mimicking the action of some neuromodulators such as
acetylcholine, may induce a transition from Up/Down state dynamics to
sustained AI states.  This transition is similar to the activation of
the brain by neuromodulators, which also can produce a transition
between slow-wave activity with Up/Down state dynamics, to the
so-called ``desynchronized'' EEG activity.  This transition naturally
occurs upon awakening from slow-wave sleep (Steriade, 2003), or can
be induced by stimulation of the ascending neuromodulatory systems
(Steriade et al, 1993a; Fig.~\ref{ppt}A), and can be mimicked by the
thalamocortical model by a reduction of the adaptation parameter $b$
(Fig.~\ref{ppt}B).  This is consistent with the action of
neuromodulators, such as acetylcholine, to block or reduce K$^+$
conductances responsible for spike-frequency adaptation (McCormick,
1992).

In this paper, there was no attempt to reproduce the correct cellular
conductance patterns of the different network states.  Experimental
measurements show that the input resistance of cortical neurons {\it
in vivo} is reduced by 3 to 5 times compared to quiescent states, in
both anesthetized animals (Contreras et al., 1996; Borg-Graham et
al., 1998; Pare et al., 1998; Destexhe and Pare, 1999) and awake
animals (Baranyi et al., 1993a, 1993b; Steriade et al., 2001;
reviewed in Destexhe et al., 2003; Destexhe, 2007).  The excitatory
and inhibitory synaptic conductances were also measured in
anesthetized (Borg-Graham et al., 1998; Rudolph et al., 2005) and
awake preparations (Rudolph et al., 2007).  Reproducing the correct
conductance state in individual neurons requires large network sizes
(El~Boustani et al., 2007; Kumar et al., 2008), and was not attempted
here.  We rather focused on the minimal size necessary to obtain AI
states in different network configurations, but obtaining states
fully consistent with experimental data will require substantial
computational resources to simulate large networks of aeIF neurons,
and constitutes a logical follow-up of this study.

It is also interesting to note that this thalamocortical model is
different than previous models of Up/Down states in cortical networks
(Timofeev et al., 2000; Compte et al., 2003; Parga and Abbott, 2007).
Previous models have considered additional mechanisms to initiate Up
states, which amounts to have some cortical neurons spontaneously
firing or external noise.  In contrast, Up/Down state dynamics arise
here entirely from self-sustained activity.  In the thalamocortical
model, the cortex is in a state generating transient dynamics (as in
Fig.~\ref{cx2000}A), while the thalamus is in an AI state.  The
system generates a transient Up state, then the activity stops,
leading to a Down state.  The activity restarts due to the firing of
TC cells which triggers a new Up state, and the cycle restarts.  This
model therefore does not generate Up/Down state dynamics in the
cortex alone, contrary to observations of Up/Down states in cortical
slices (Sanchez-Vives and McCormick, 2000; Cossart et al., 2003) or
in cortical organotypic cultures (Plenz and Aertsen, 1996).  It
nevertheless accounts for the fact that the thalamus ignites the Up
states in the intact thalamocortical system {\it in vivo} (Contreras
and Steriade, 1995; Grenier et al., 1998).

Similar Up/Down state dynamics were also observed in a more
sophisticated model of the cortex allowing layered connectivity, in
which a small sub-network played the role of AI state generator,
while the rest of the network was passive and generated transient
dynamics, as in Fig.~\ref{cx2000}A.  The interconnections between two
such networks generated Up/Down state dynamics very similar to the
thalamocortical model, except that in this case, the activity was
entirely generated within the cortex, in agreement with cortical
slice experiments.  The sub-network displaying self-sustained AI
states (Layer~B in Fig.~\ref{2layers}) may represent Layer~V
networks, which were reported to initiate the activity during Up
states in cortical slices (Sanchez-Vives and McCormick, 2000).  Also
similar to the thalamocortical model, this two-layer model of the
cortex generated Up/Down state dynamics entirely from the internal
dynamics of the system, without the requirement of spontaneously
firing cells, or additional noise, as in previous models.  The
bistability of this system was illustrated by the coexistence of a
stable resting state and Up/Down state dynamics (Fig.~\ref{2layers},
arrow).  Note that here, the specific choice of networks with LTS
cells (layer~B) should not be considered as a prediction, as similar
dynamics should also be observed if Layer~B was a network of RS and
FS cells displaying AI states (as in Fig.~\ref{cx2000}B).  The
essential prediction here is that a network with self-sustained
activity, combined with another network displaying spike-frequency
adaptation, is a possible generator of self-sustained Up/Down state
dynamics.

Finally, the present model reports a variety of AI states in
different network configurations, but it is important to note that AI
states may be transient in nature.  Transient chaotic regimes have
been extensively studied (Crutchfield and Kaneko, 1988; Zillmer et
al., 2006; Cessac, 2008; Cessac and Vi\'eville, 2009; reviewed in
T\'el and Lai, 2008).  The lifetime of these transient regimes is
known to increase exponentially with network size (Crutchfield and
Kaneko, 1988; Kumar et al., 2008; T\'el and Lai, 2008; El Boustani
and Destexhe, 2009).  For large networks, of the order of thousand or
more neurons, this lifetime can reach considerable times (beyond any
reasonable simulation time; see Kumar et al., 2008).  It is thus
difficult to formally establish the transient nature of such states
with numerical simulations.  However, although AI states may not
represent a stable attractor of the system, they are nevertheless
relevant because they represent the dynamics of the system within any
reasonably accessible time period.

In conclusion, we have shown here what should be considered as the
beginning of an exploration of the collective dynamics of populations
of neurons endowed with complex intrinsic properties as given by the
aeIF model.  The presence of such properties has important effects on
network behavior, and in particular affects the genesis of AI state. 
AI states are not possible in small networks without LTS properties,
which certainly constitutes an interesting prediction.  More
generally, future studies should investigate intrinsic properties as
a continuum rather than distinct neuronal classes.  Such a continuum
should be possible to define using IF models such as the aeIF model
used here.


\subsection*{Acknowledgments}

Thanks to Sami El~Boustani for comments on the manuscript.  Research
supported by the Centre National de la Recherche Scientifique (CNRS,
France), Agence Nationale de la Recherche (ANR, France) and the
Future and Emerging Technologies program (FET, European Union; FACETS
project).  Additional information is available at
\url{http://cns.iaf.cnrs-gif.fr}


%
%
%
%

\section*{References}

\begin{description}

\bibitem{Avendano88} Avenda\~no C, Rausell E, Perez-Aguilar D, and
Isorna S. (1988)  Organization of the association cortical afferent
connections of area 5: a retrograde tracer study in the cat. J. 
Comp. Neurol. 278: 1-33.
 
\bibitem{Avendano85} Avenda\~no C, Rausell E and Reinoso-Suarez F. 
(1985) Thalamic projections to areas 5a and 5b of the parietal cortex
in the cat: a retrograde horseradish peroxidase study.  J. Neurosci. 
5: 1446-1470.

\bibitem{BarWoo93a} Baranyi A., Szente M.B., Woody C.D. (1993a)
Electrophysiological characterization of different types of neurons
recorded in vivo in the motor cortex of the cat. I. patterns of
firing activity and synaptic responses. J. Neurophysiol. 69:
1850-1864.

\bibitem{BarWoo93b} Baranyi A., Szente M.B., Woody C.D. (1993b)
Electrophysiological characterization of different types of neurons
recorded {\it in vivo} in the motor cortex of the cat. II. Membrane
parameters, action potentials, current-induced voltage responses and
electrotonic structures. J. Neurophysiol. 69: 1865-1879.

\bibitem{Binzegger2004} Binzegger T, Douglas RJ and Martin, KAC. 
(2004) A quantitative map of the circuit of cat primary visual
cortex. J. Neurosci. 24: 8441-8453.

\bibitem{BorFre98} Borg-Graham L.J., Monier C., Fr\'egnac Y. (1998)
Visual input evokes transient and strong shunting inhibition in
visual cortical neurons.  Nature 393: 369-373.

\bibitem{Bourassa95} Bourassa J and Desch\^enes M. (1995)
Corticothalamic projections from the primary visual cortex in rats: a
single fiber study using biocytin as an anterograde tracer. 
Neuroscience 66: 253-263.

\bibitem{Braitenberg98} Braitenberg, V. and Sch\"uz, A. (1998) {\it
Cortex: statistics and geometry of neuronal connectivity} (2nd
edition).  Springer-Verlag, Berlin.

\bibitem{Brette2005} Brette R and Gerstner W. (2005) Adaptive
exponential integrate-and-fire model as an effective description of
neuronal activity. J. Neurophysiol. 94: 3637-3642.

\bibitem{Brunel2000} Brunel N. (2000)  Dynamics of sparsely connected
networks of excitatory and inhibitory spiking neurons. J. Comput.
Neurosci. 8: 183-208.  

\bibitem{Cessac2008} Cessac, B. (2008) A discrete time neural network
model with spiking neurons. Rigorous results on the spontaneous
dynamics.  J. Math. Biol. 56: 311-345.

\bibitem{Cessac-Vieville2008} Cessac, B. and Vi\'eville, T.  (2009)
On dynamics of integrate-and-fire neural networks with conductance
based synapses.  Front. Comput. Neurosci. 3: 1.

\bibitem{Crutchfield88} Crutchfield, J.P. and Kaneko, K. (1988) Are
attractors relevant to turbulence?  Phys. Rev. Lett. 60: 2715-2718.

\bibitem{Compte2003} Compte, A., Sanchez-Vives, M.V., McCormick, D.A.
and Wang, X.J. (2003) Cellular and network mechanisms of slow
oscillatory activity ($<$1~Hz) and wave propagations in a cortical
network model.  J. Neurophysiol. 89: 2707-2725.

\bibitem{Connors90} Connors, B.W. and Gutnick, M.J. (1990) Intrinsic
Firing patterns of diverse neocortical neurons. Trends Neurosci. 13:
99-104.

\bibitem{Contreras95} Contreras D and Steriade M (1995) Cellular
basis of EEG slow rhythms: a study of dynamic corticothalamic
relationships. J. Neurosci. 15: 604-622.

\bibitem{ConSte96} Contreras D., Timofeev I., Steriade M. (1996)
Mechanisms of long lasting hyperpolarizations underlying slow sleep
oscillations in cat corticothalamic networks. J. Physiol. 494:
251-264.

\bibitem{Cossart2003} Cossart R, Aronov D, Yuste R. (2003) Attractor
dynamics of network UP states in the neocortex. Nature 423: 283-238.

\bibitem{delaPena96} de la Pe\~na E and Geijo-Barrientos E. (1996) 
Laminar organization, morphology and physiological properties of
pyramidal neurons that have the low-threshold calcium current in the
guinea-pig frontal cortex.  J. Neurosci. 16: 5301-5311.

\bibitem{Destexhe2007} Destexhe, A. (2007) High-conductance state. 
Scholarpedia 2(11): 1341 \\ {\small
\url{http://www.scholarpedia.org/article/High-Conductance\_State}}

\bibitem{Destexhe-Contreras2006} Destexhe A and Contreras D. (2006)
Neuronal computations with stochastic network states. Science 314:
85-90.

\bibitem{DesPar99} Destexhe A and Par\'e D. (1999) Impact of network
activity on the integrative properties of neocortical pyramidal
neurons in vivo. J. Neurophysiol. 81: 1531-1547.

\bibitem{JNP98} Destexhe A, Contreras D and Steriade M. (1998)
Mechanisms underlying the synchronizing action of corticothalamic
feedback through inhibition of thalamic relay cells. J. 
Neurophysiol. 79: 999-1016.

\bibitem{CxSW} Destexhe A, Contreras D and Steriade M. (2001) LTS
cells in cerebral cortex and their role in generating spike-and-wave
oscillations.   Neurocomputing 38: 555-563.

\bibitem{TINS2007} Destexhe A., Hughes SW, Rudolph M, Crunelli V.
(2007)  Are corticothalamic `up' states fragments of wakefulness?
Trends Neurosci. 30: 334-342.

\bibitem{DP99} Destexhe A, Par\'e D. (1999) Impact of network
activity on the integrative properties of neocortical pyramidal
neurons in vivo. J.  Neurophysiol. 81: 1531-1547.

\bibitem{Dest2003} Destexhe A, Rudolph, M, Par\'e D (2003) The
high-conductance state of neocortical neurons in vivo.  Nature
Reviews Neurosci. 4: 739-751.

\bibitem{PhysiolRev2003} Destexhe A, Sejnowski TJ. (2003)
Interactions between membrane conductances underlying thalamocortical
slow-wave oscillations.  Physiol. Reviews 83: 1401-1453.

\bibitem{Master2008} El Boustani, S. and Destexhe, A. (2009)  A
master equation formalism for macroscopic modeling of asynchronous
irregular activity states. Neural Comput. 21: 46-100.

\bibitem{Sami2007} El Boustani S, Pospischil M, Rudolph-Lilith M,
Destexhe A.  (2007) Activated cortical states: experiments, analyses
and models.  J. Physiol. Paris 101: 99-109.

\bibitem{FitzGibbon95} FitzGibbon T, Tevah LV and Jervie-Sefton A. 
(1995) Connections between the reticular nucleus of the thalamus and
pulvinar-lateralis posterior complex: a WGA-HRP study.  J. Comp. 
Neurol. 363: 489-504.

\bibitem{Fourcaud2003} Fourcaud-Trocme N, Hansel D, van Vreeswijk C
and Brunel N. (2003) How spike generation mechanisms determine the
neuronal response to fluctuating inputs. J. Neurosci. 23:
11628-11640.  

\bibitem{Freund89} Freund TF, Martin KA, Soltesz I, Somogyi P and
Whitteridge D. (1989)  Arborisation pattern and postsynaptic targets
of physiologically identified thalamocortical afferents in striate
cortex of the macaque monkey. J. Comp. Neurol. 289: 315-336.

\bibitem{Grenier98} Grenier F, Timofeev I and Steriade M.  (1998)
Leading role of thalamic over cortical neurons during postinhibitory
rebound excitation.  Proc. Natl. Acad. Sci. USA 95: 13929-13934.

\bibitem{Hines97} Hines, M.L. and Carnevale, N.T. (1997). The Neuron
simulation environment. Neural Computation 9: 1179-1209.

\bibitem{Izhikevich2004} Izhikevich EM. (2004) Which model to use for
cortical spiking neurons? IEEE Trans. Neural Networks.  15: 1063-1070.

\bibitem{Jones:Thalamus} Jones, E.G. (1985)  {\it The Thalamus}. 
Plenum Press, New York.

\bibitem{Kim97} Kim U, Sanches-Vives MV and McCormick DA. (1997)
Functional dynamics of GABAergic inhibition in the thalamus.  Science
278: 130-134.

\bibitem{Kumar2008} Kumar A, Schrader S, Aertsen A, Rotter S. (2008) 
The high-conductance state of cortical networks. Neural Comput. 20:
1-43.

\bibitem{Landry81} Landry P and Desch\^enes M. (1981) Intracortical
arborizations and receptive fields of identified ventrobasal
thalamocortical afferents to the primary somatic sensory cortex in
the cat. J. Comp. Neurol. 199: 345-371.

\bibitem{Lee2006} Lee AK, Manns ID, Sakmann B and Brecht M.  (2006)
Whole-cell recordings in freely moving rats. Neuron 51: 399-407.

\bibitem{Llinas88} Llin\'as RR. (1988) The intrinsic
electrophysiological properties of mammalian neurons: a new insight
into CNS function. Science 242: 1654-1664.

\bibitem{Matsumara88} Matsumura M, Cope T, Fetz EE. (1988) Sustained
excitatory synaptic input to motor cortex neurons in awake animals
revealed by intracellular recording of membrane potentials. Exp. 
Brain Res. 70: 463-469.

\bibitem{McCormick92} McCormick DA. (1992)  Neurotransmitter actions
in the thalamus and cerebral cortex and their role in neuromodulation
of thalamocortical activity.  Prog. Neurobiol. 39: 337-388.

\bibitem{Minderhoud71} Minderhoud JM. (1971) An anatomical study of
the efferent connections of the thalamic reticular nucleus.  Exp.
Brain Res. 112: 435-446.

\bibitem{Muller2007} Muller E, Buesing L, Schemmel J, Meier K. (2007)
Spike-frequency adapting neural ensembles: beyond mean adaptation and
renewal theories.  Neural Computation 19: 2958-3010.  

\bibitem{Pare98} Par\'e D, Shink E, Gaudreau H, Destexhe A, Lang EJ.
(1998) Impact of spontaneous synaptic activity on the resting
properties of cat neocortical neurons in vivo. J. Neurophysiol. 79:
1450-1460.

\bibitem{Parga2007} Parga N and Abbott LF. (2007) Network model of
spontaneous activity exhibiting synchronous transitions between up
and down States.  Front. Neurosci. 1: 57-66.

\bibitem{Plenz96} Plenz, D., and Aertsen, A. (1996) Neural dynamics
in cortex-striatum co-cultures II - spatiotemporal characteristics of
neuronal activity.  Neuroscience 70: 893-924.

\bibitem{Lausanne2008} Pospischil, M., Toledo-Rodriguez, M., Monier,
C., Piwkowska, Z., Bal, T., Fr\'egnac, Y., Markram, H. and Destexhe,
A. (2008)  Minimal Hodgkin-Huxley type models for different classes
of cortical and thalamic neurons. Biol. Cybernetics 99: 427-441.

\bibitem{Raussell95} Rausell E and Jones EG. (1995)  Extent of
intracortical arborization of thalamocortical axons as a determinant
of representational plasticity in monkey somatic sensory cortex.  J.
Neurosci. 15: 4270-4288.

\bibitem{Robertson81} Robertson RT and Cunningham TJ. (1981)
Organization of corticothalamic projections from parietal cortex in
cat. J.  Comp.  Neurol. 199: 569-585.

\bibitem{PPT} Rudolph, M., Pelletier, J-G., Par\'e, D. and Destexhe,
A.  (2005)  Characterization of synaptic conductances and integrative
properties during electrically-induced EEG-activated states in
neocortical neurons in vivo. J. Neurophysiol. 94: 2805-2821.

\bibitem{Rud2007} Rudolph, M., Pospischil, M., Timofeev, I. and
Destexhe, A.  (2007) Inhibition determines membrane potential
dynamics and controls action potential generation in awake and
sleeping cat cortex. J.  Neurosci. 27: 5280-5290.

\bibitem{Sanchez2000} Sanchez-Vives, MV and McCormick, DA. (2000)
Cellular and network mechanisms of rhythmic recurrent activity in
neocortex. Nat. Neurosci. 10: 1027-1034.

\bibitem{Sherman2001} Sherman, S. M. \& Guillery, R. W. (2001) {\it
Exploring the Thalamus.} Academic Press, New York.

\bibitem{Smith2000} Smith GD, Cox CL, Sherman M and Rinzel J. (2000).
Fourier analysis of sinusoidally driven thalamocortical relay neurons
and a minimal integrate-and-fire-or-burst model. J. Neurophysiol. 83:
588-610.

\bibitem{Ster70} Steriade M. (1970) Ascending control of thalamic and
cortical responsiveness. Int. Rev. Neurobiol. 12: 87-144.

\bibitem{Ster2001} Steriade, M. (2001)  Impact of network activities
on neuronal properties in corticothalamic systems.  J. Neurophysiol.
86: 1-39.

\bibitem{Ster2003} Steriade, M. (2003)  {\it Neuronal Substrates of
Sleep and Epilepsy}.  Cambridge University Press, Cambridge, UK.

\bibitem{Ster93a} Steriade M, Amzica F, Nunez A. (1993a) Cholinergic
and noradrenergic modulation of the slow ($\sim$0.3 Hz) oscillation
in neocortical cells.  J Neurophysiol 70: 1384-1400.

\bibitem{Steriade85} Steriade M, Desch\^enes M, Domich L and Mulle C.
(1985) Abolition of spindle oscillations in thalamic neurons
disconnected from nucleus reticularis thalami. J. Neurophysiol.  54:
1473-1497.

\bibitem{Ster90} Steriade, M., McCarley, R.W. (1990)  {\it Brainstem
Control of Wakefulness and Sleep}, Plenum Press, New York.

\bibitem{Steriade:slow2} Steriade M, Nunez A and Amzica F.  (1993b)
Intracellular analysis of relations between the slow ($<$~1~$Hz$)
neocortical oscillation and other sleep rhythms of the
electroencephalogram. J. Neurosci. 13: 3266-3283.

\bibitem{Steriade2001} Steriade M, Timofeev I, Grenier F. (2001)
Natural waking and sleep states: a view from inside neocortical
neurons.  J.  Neurophysiol. 85: 1969-1985.

\bibitem{Tel2008} T\'el, T. and Lai, Y.-C. (2008) Chaotic transients
in spatially extended systems.  Physics Rep. 460: 245-275.

\bibitem{Timofeev2000} Timofeev I, Grenier F, Bazhenov M, Sejnowski
TJ and Steriade M. (2000)  Origin of slow cortical oscillations in
deafferented cortical slabs. Cereb. Cortex 10: 1185-1199.

\bibitem{Thomson2003} Thomson AM and Bannister AP. (2003)
Interlaminar connections in the neocortex. Cereb. Cortex 13: 5-14.

\bibitem{Updyke81} Updyke BV. (1981)  Projections from visual areas
of the middle suprasylvian sulcus onto the lateral posterior complex
and adjacent thalamic nuclei in cat. J. Comp. Neurol.  201: 477-506.

\bibitem{Vogels2005} Vogels TP and Abbott LF. (2005) Signal
propagation and logic gating in networks of integrate-and-fire
neurons. J.  Neurosci. 25: 10786-10795.

\bibitem{McCormick:spindles} von Krosigk M, Bal T and McCormick, DA. 
(1993) Cellular mechanisms of a synchronized oscillation in the
thalamus. Science 261: 361-364.

\bibitem{White86} White EL. (1986) Termination of thalamic afferents
in the cerebral cortex.  In: {\it Cerebral Cortex} (Jones EG, Peters
A, eds) Vol.~5, pp 271-289.  New York: Plenum Press.

\bibitem{White-Hersch82} White EL and Hersch SM. (1982) A
quantitative study of thalamocortical and other synapses involving
the apical dendrites of corticothalamic cells in mouse SmI cortex. 
J.  Neurocytol. 11: 137-157.

\bibitem{Xiang98} Xiang, Z., Huguenard, J.R. and Prince, D.A. (1998)
Cholinergic switching within neocortical inhibitory networks. Science
281: 985-988.

\bibitem{Zillmer2006} Zillmer R, Livi R, Politi A and Torcini A.
(2006).  Desynchronization in diluted neural networks.  Phys. Rev. E.
74: 036203.

\end{description}

\end{document}